\pdfoutput=1

\documentclass[12pt,a4paper]{article}

\usepackage{ifthen} 
\newboolean{pdflatex}
\setboolean{pdflatex}{true} 

\newboolean{articletitles}
\setboolean{articletitles}{true} 

\newboolean{uprightparticles}
\setboolean{uprightparticles}{false} 

\newboolean{inbibliography}
\setboolean{inbibliography}{false} 

\def\paperauthors{LHCb collaboration} 
\def\paperasciititle{Measurements of charm mixing and CP violation using D0 -> K+/- pi-/+ decays} 
\def\papertitle{Measurements of charm mixing and $C\!P$ violation using $D^0 \to K^\pm \pi^\mp$ decays} 
\def\paperkeywords{{High Energy Physics}, {LHCb}, {Charm quark}, {CP violation}, {Neutral-meson mixing}} 
\def\papercopyright{\the\year\ CERN for the benefit of the LHCb collaboration} 
\def\paperlicence{CC BY 4.0 licence}
\def\paperlicenceurl{https://creativecommons.org/licenses/by/4.0/}

\usepackage[top=1in, bottom=1.25in, left=1in, right=1in]{geometry}

\usepackage{microtype}
\usepackage{lineno}  
\usepackage{xspace} 
\usepackage{caption} 

\usepackage{graphicx}  
\usepackage{color}
\usepackage{colortbl}
\graphicspath{{./figs/}} 

\usepackage{amsmath} 
\usepackage{amssymb}
\usepackage{amsfonts}
\usepackage{upgreek} 

\newcommand*\patchAmsMathEnvironmentForLineno[1]{%
\expandafter\let\csname old#1\expandafter\endcsname\csname #1\endcsname
\expandafter\let\csname oldend#1\expandafter\endcsname\csname
end#1\endcsname
 \renewenvironment{#1}%
   {\linenomath\csname old#1\endcsname}%
   {\csname oldend#1\endcsname\endlinenomath}%
}
\newcommand*\patchBothAmsMathEnvironmentsForLineno[1]{%
  \patchAmsMathEnvironmentForLineno{#1}%
  \patchAmsMathEnvironmentForLineno{#1*}%
}
\AtBeginDocument{%
\patchBothAmsMathEnvironmentsForLineno{equation}%
\patchBothAmsMathEnvironmentsForLineno{align}%
\patchBothAmsMathEnvironmentsForLineno{flalign}%
\patchBothAmsMathEnvironmentsForLineno{alignat}%
\patchBothAmsMathEnvironmentsForLineno{gather}%
\patchBothAmsMathEnvironmentsForLineno{multline}%
\patchBothAmsMathEnvironmentsForLineno{eqnarray}%
}

\usepackage{hyperxmp}
\usepackage[pdftex,
            pdfauthor={\paperauthors},
            pdftitle={\paperasciititle},
            pdfkeywords={\paperkeywords},
            pdfcopyright={Copyright (C) \papercopyright},
            pdflicenseurl={\paperlicenceurl}]{hyperref}
\usepackage[all]{hypcap} 

\usepackage{xspace} 
\usepackage{upgreek}

\def\lhcb {\mbox{LHCb}\xspace}

\def\MagUp {\mbox{\em Mag\kern -0.05em Up}\xspace}

\ifthenelse{\boolean{uprightparticles}}%
{

 \def\Pmu         {\ensuremath{\upmu}\xspace}

 \def\Ppi         {\ensuremath{\uppi}\xspace}

 \def\PDelta      {\ensuremath{\Delta}\xspace}                 
 \def\PXi      {\ensuremath{\Xi}\xspace}                 
 \def\PLambda      {\ensuremath{\Lambda}\xspace}                 
 \def\PSigma      {\ensuremath{\Sigma}\xspace}                 
 \def\POmega      {\ensuremath{\Omega}\xspace}                 
 \def\PUpsilon      {\ensuremath{\Upsilon}\xspace}                 
 
 \def\PB      {\ensuremath{\mathrm{B}}\xspace}                 
                  
 \def\PD      {\ensuremath{\mathrm{D}}\xspace}

 \def\PK      {\ensuremath{\mathrm{K}}\xspace}

 \def\Pb      {\ensuremath{\mathrm{b}}\xspace}                 
 \def\Pc      {\ensuremath{\mathrm{c}}\xspace}

 \def\Pi      {\ensuremath{\mathrm{i}}\xspace}

 \def\Pp      {\ensuremath{\mathrm{p}}\xspace}

 \def\Ps      {\ensuremath{\mathrm{s}}\xspace}

}
{

 \def\Pmu         {\ensuremath{\mu}\xspace}

 \def\Ppi         {\ensuremath{\pi}\xspace}

 \mathchardef\PDelta="7101
 \mathchardef\PXi="7104
 \mathchardef\PLambda="7103
 \mathchardef\PSigma="7106
 \mathchardef\POmega="710A
 \mathchardef\PUpsilon="7107
                  
 \def\PB      {\ensuremath{B}\xspace}                 
                  
 \def\PD      {\ensuremath{D}\xspace}

 \def\PK      {\ensuremath{K}\xspace}

 \def\Pb      {\ensuremath{b}\xspace}                 
 \def\Pc      {\ensuremath{c}\xspace}

 \def\Pi      {\ensuremath{i}\xspace}

 \def\Pp      {\ensuremath{p}\xspace}

 \def\Ps      {\ensuremath{s}\xspace}

}

\makeatletter
\ifcase \@ptsize \relax
  \newcommand{\miniscule}{\@setfontsize\miniscule{4}{5}}
\or
  \newcommand{\miniscule}{\@setfontsize\miniscule{5}{6}}
\or
  \newcommand{\miniscule}{\@setfontsize\miniscule{5}{6}}
\fi
\makeatother

\DeclareRobustCommand{\optbar}[1]{\shortstack{{\miniscule (\rule[.5ex]{1.25em}{.18mm})}
  \\ [-.7ex] $#1$}}

\def\mup        {{\ensuremath{\Pmu^+}}\xspace}
\def\mun        {{\ensuremath{\Pmu^-}}\xspace} 

\def\squark    {{\ensuremath{\Ps}}\xspace}

\def\cquark    {{\ensuremath{\Pc}}\xspace}

\def\bquark    {{\ensuremath{\Pb}}\xspace}

\def\pion   {{\ensuremath{\Ppi}}\xspace}

\def\pip    {{\ensuremath{\pion^+}}\xspace}
\def\pim    {{\ensuremath{\pion^-}}\xspace}

\def\pimp   {{\ensuremath{\pion^\mp}}\xspace}

\def\kaon    {{\ensuremath{\PK}}\xspace}
  \def\Kbar    {{\kern 0.2em\overline{\kern -0.2em \PK}{}}\xspace}

\def\KorKbar    {\kern 0.18em\optbar{\kern -0.18em K}{}\xspace}
\def\Kz      {{\ensuremath{\kaon^0}}\xspace}
\def\Kzb     {{\ensuremath{\Kbar{}^0}}\xspace}
\def\Kp      {{\ensuremath{\kaon^+}}\xspace}
\def\Km      {{\ensuremath{\kaon^-}}\xspace}
\def\Kpm     {{\ensuremath{\kaon^\pm}}\xspace}

\def\KS      {{\ensuremath{\kaon^0_{\mathrm{ \scriptscriptstyle S}}}}\xspace}

  \def\Dbar    {{\kern 0.2em\overline{\kern -0.2em \PD}{}}\xspace}
\def\D       {{\ensuremath{\PD}}\xspace}

\def\DorDbar    {\kern 0.18em\optbar{\kern -0.18em D}{}\xspace}
\def\Dz      {{\ensuremath{\D^0}}\xspace}
\def\Dzb     {{\ensuremath{\Dbar{}^0}}\xspace}
\def\Dp      {{\ensuremath{\D^+}}\xspace}
\def\Dm      {{\ensuremath{\D^-}}\xspace}
\def\Dpm     {{\ensuremath{\D^\pm}}\xspace}

\def\Dstarp  {{\ensuremath{\D^{*+}}}\xspace}
\def\Dstarm  {{\ensuremath{\D^{*-}}}\xspace}

\def\B       {{\ensuremath{\PB}}\xspace}
\def\Bbar    {{\ensuremath{\kern 0.18em\overline{\kern -0.18em \PB}{}}}\xspace}

\def\BorBbar    {\kern 0.18em\optbar{\kern -0.18em B}{}\xspace}

\def\Bs      {{\ensuremath{\B^0_\squark}}\xspace}
\def\Bsb     {{\ensuremath{\Bbar{}^0_\squark}}\xspace}

  \def\Y#1S{\ensuremath{\PUpsilon{(#1S)}}\xspace}

\def\proton      {{\ensuremath{\Pp}}\xspace}

\def\Lbar        {{\ensuremath{\kern 0.1em\overline{\kern -0.1em\PLambda}}}\xspace}
\def\LorLbar    {\kern 0.18em\optbar{\kern -0.18em \PLambda}{}\xspace}

\newcommand{\decay}[2]{\ensuremath{#1\!\to #2}\xspace}         

\def\to                 {\ensuremath{\rightarrow}\xspace}

\def\CP                {{\ensuremath{C\!P}}\xspace}

\def\AT#1     {\ensuremath{A_{\mathrm{T}}^{#1}}\xspace}           

\def\C#1      {\ensuremath{\mathcal{C}_{#1}}\xspace}                       
\def\Cp#1     {\ensuremath{\mathcal{C}_{#1}^{'}}\xspace}                    
\def\Ceff#1   {\ensuremath{\mathcal{C}_{#1}^{\mathrm{(eff)}}}\xspace}        
\def\Cpeff#1  {\ensuremath{\mathcal{C}_{#1}^{'\mathrm{(eff)}}}\xspace}       
\def\Ope#1    {\ensuremath{\mathcal{O}_{#1}}\xspace}                       
\def\Opep#1   {\ensuremath{\mathcal{O}_{#1}^{'}}\xspace}                    

\newcommand{\bra}[1]{\ensuremath{\langle #1|}}             
\newcommand{\ket}[1]{\ensuremath{|#1\rangle}}              

\newcommand{\tev}{\ifthenelse{\boolean{inbibliography}}{\ensuremath{~T\kern -0.05em eV}\xspace}{\ensuremath{\mathrm{\,Te\kern -0.1em V}}}\xspace}
\newcommand{\gev}{\ensuremath{\mathrm{\,Ge\kern -0.1em V}}\xspace}
\newcommand{\mev}{\ensuremath{\mathrm{\,Me\kern -0.1em V}}\xspace}
\newcommand{\kev}{\ensuremath{\mathrm{\,ke\kern -0.1em V}}\xspace}
\newcommand{\ev}{\ensuremath{\mathrm{\,e\kern -0.1em V}}\xspace}
\newcommand{\gevc}{\ensuremath{{\mathrm{\,Ge\kern -0.1em V\!/}c}}\xspace}
\newcommand{\mevc}{\ensuremath{{\mathrm{\,Me\kern -0.1em V\!/}c}}\xspace}
\newcommand{\gevcc}{\ensuremath{{\mathrm{\,Ge\kern -0.1em V\!/}c^2}}\xspace}
\newcommand{\gevgevcccc}{\ensuremath{{\mathrm{\,Ge\kern -0.1em V^2\!/}c^4}}\xspace}
\newcommand{\mevcc}{\ensuremath{{\mathrm{\,Me\kern -0.1em V\!/}c^2}}\xspace}

\def\mum  {\ensuremath{{\,\upmu\mathrm{m}}}\xspace}

\def\invfb   {\ensuremath{\mbox{\,fb}^{-1}}\xspace}

\newcommand{\chisq}{\ensuremath{\chi^2}\xspace}
\newcommand{\chisqndf}{\ensuremath{\chi^2/\mathrm{ndf}}\xspace}
\newcommand{\chisqip}{\ensuremath{\chi^2_{\text{IP}}}\xspace}

\def\gsim{{~\raise.15em\hbox{$>$}\kern-.85em
          \lower.35em\hbox{$\sim$}~}\xspace}
\def\lsim{{~\raise.15em\hbox{$<$}\kern-.85em
          \lower.35em\hbox{$\sim$}~}\xspace}

\def\sqs   {\ensuremath{\protect\sqrt{s}}\xspace}

\def\ptot       {\mbox{$p$}\xspace}
\def\pt         {\mbox{$p_{\mathrm{ T}}$}\xspace}

\def\tell1  {TELL1\xspace}
\def\ukl1   {UKL1\xspace}

\newcommand{\ie}{\mbox{\itshape i.e.}\xspace}

\usepackage{cite} 
\usepackage{mciteplus}
\usepackage{siunitx}

\begin{document}

\renewcommand{\thefootnote}{\fnsymbol{footnote}}
\setcounter{footnote}{1}

\def\dbar{\overline{D}{}^{\,0}}
\def\dzket{\left| D^0\right.\rangle}
\def\dzbarket{\left| \dbar\right.\rangle}
\def\dzkett{\left| D^0(t)\right.\rangle}
\def\dzbarkett{\left| \dbar(t)\right.\rangle}


\begin{titlepage}
\pagenumbering{roman}

\vspace*{-1.5cm}
\centerline{\large EUROPEAN ORGANIZATION FOR NUCLEAR RESEARCH (CERN)}
\vspace*{1.5cm}
\noindent
\begin{tabular*}{\linewidth}{lc@{\extracolsep{\fill}}r@{\extracolsep{0pt}}}
\ifthenelse{\boolean{pdflatex}}
{\vspace*{-2.7cm}\mbox{\!\!\!\includegraphics[width=.14\textwidth]{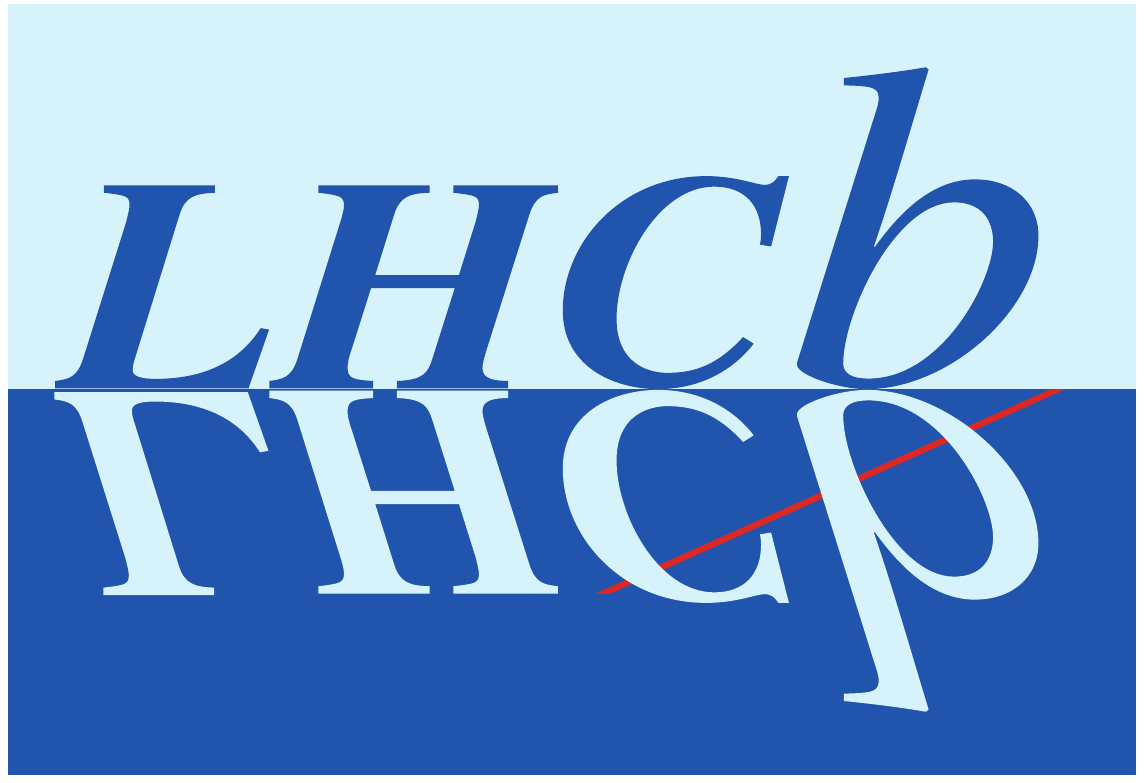}} & &}%
{\vspace*{-1.2cm}\mbox{\!\!\!\includegraphics[width=.12\textwidth]{lhcb-logo.eps}} & &}%
\\
 & & CERN-EP-2016-280 \\  
 & & LHCb-PAPER-2016-033 \\  
 & & \today \\ 
 & & \\
\end{tabular*}

\vspace*{4.0cm}

{\normalfont\bfseries\boldmath\huge
\begin{center}
  \papertitle
\end{center}
}

\vspace*{5mm}

\begin{center}
\paperauthors 
\footnote{Authors are listed at the end of this paper.}
\end{center}

\vspace*{5mm}
\begin{abstract}
  \noindent
  Measurements of charm mixing and \CP violation parameters
  from the decay-time-dependent ratio of $ D^0 \to K^+ \pi^- $
  to $ D^0 \to K^- \pi^+ $ decay rates and the charge-conjugate ratio
  are reported.
  The analysis uses
  $\overline{B}\to \Dstarp \mun X$, and charge-conjugate decays, where $\Dstarp\to \Dz \pip$, and
  $\Dz\to K^{\mp} \pi^{\pm}$. The
  $pp$ collision data are recorded by the \lhcb experiment at 
  center-of-mass energies \sqs = 7 and 8 \tev,
  corresponding to an integrated luminosity of 3\invfb.
  The data are analyzed under three hypotheses:
  (i) mixing assuming \CP symmetry, (ii) mixing assuming no direct
  \CP violation in the Cabibbo-favored or
  doubly Cabibbo-suppressed decay amplitudes, and (iii) mixing allowing  
  either direct \CP violation and/or \CP violation in the superpositions of flavor eigenstates defining the mass eigenstates. 
  The data are also combined with those from a previous 
  \lhcb study of $\Dz\to K \pi$ decays
  from a disjoint set of $ D^{*+} $ candidates produced directly in $pp$ collisions. In all cases, the data are consistent
  with the hypothesis of \CP symmetry.
  
\end{abstract}

\vspace*{10mm}

\begin{center}
  Published in \href{https://doi.org/10.1103/PhysRevD.95.052004}{Phys.~Rev.~{\bf D95}, 052004 (2017)}\\
  Errata published in
  \href{https://doi.org/10.1103/PhysRevD.96.099907}{Phys.~Rev.~{\bf D96}, 099907 (2017)} and
  \href{https://doi.org/10.1103/PhysRevD.111.039901}{Phys.~Rev.~{\bf D111}, 039901 (2025)}\footnote{The contents of the errata are reflected in this manuscript.}
\end{center}
\vspace*{0.2cm}
\vspace{\fill}
{\footnotesize
\centerline{\copyright~\papercopyright. \href{\paperlicenceurl}{\paperlicence}.}}
\vspace*{2mm}

\end{titlepage}


\newpage
\setcounter{page}{2}
\mbox{~}

\cleardoublepage


\renewcommand{\thefootnote}{\arabic{footnote}}
\setcounter{footnote}{0}



\pagestyle{plain} 
\setcounter{page}{1}
\pagenumbering{arabic}


%

\section{Introduction}
\label{sec:Introduction}

The oscillation of \Dz mesons into \Dzb mesons,
and  vice versa, is a manifestation
of the fact that the flavor and mass 
eigenstates of the neutral charm meson system differ.
Such oscillations are also referred to as 
mixing. 
Charge-parity violation ($CPV$) in the superpositions of flavor eigenstates
defining the mass eigenstates can lead to
different mixing rates for
\Dz into \Dzb and \Dzb into \Dz.
The LHCb experiment has previously reported measurements of 
mixing and $\CP$
violation parameters from studies of $\Dstarp\to \Dz \pi_s^+,\,\Dz \to K^\pm \pi^\mp $
decays,
where the where the \Dstarp meson is produced directly in $pp$
collisions~\cite{LHCb-PAPER-2013-053}. In this sample, referred 
to as ``prompt,'' the flavor of the \Dz mesons at the production is determined 
by the charge of the slow pion $\pi_s^+$ from the strong decay of the $\Dstarp$ meson. 
In this paper we extend the study using $ \Dz $ mesons 
produced in $ \overline{B} \to D^{*+}\mun X$, 
$D^{*+} \to D^0 \pi^+_s$, $D^0 \to K^{\pm} \pi^{\mp} $ and
charge-conjugate decays\footnote{Except when otherwise explicitly stated,
charge-conjugate processes are implied.}, using $pp$ 
collision data recorded by the \lhcb experiment at 
center-of-mass energies \sqs = 7 and 8 \tev,
corresponding to an integrated luminosity of 3\invfb.
In this case, the flavor of the $\Dz$ at production is tagged twice, once by
the charge of the muon and once by the 
opposite charge of the slow
pion  $ \pi_s^+ $ produced in the $ D^{*+} $ decay, leading to very pure
samples.
The doubly tagged (DT)
$ \overline{B} \to D^{*+} \mun X $ candidates selected by the trigger
are essentially
unbiased with respect to the $ \Dz $ decay time, while
those in the prompt sample are
selected 
by the trigger with a bias towards
higher decay times.
As a result, the DT analysis allows for better measurements
at lower decay times.
In this paper, we first report the results of a mixing and $CPV$
analysis using the DT sample,
and then report the results of simultaneous fits to the DT and prompt samples.

\section{Theoretical Framework}
\label{sec:theory}
The physical eigenstates of the neutral $D$ system,
which have well-defined masses and lifetimes, can be written as 
linear combinations of the flavor eigenstates, which
have well-defined quark content: 
$\ket{D_{1,2}}=p\ket{\Dz}\pm q\ket{\Dzb}$. We follow the convention 
$(\CP) \ket{\Dz} =  -\ket{\Dzb}$~\cite{HFAG}.
The coefficients $p$ and $q$ are complex numbers, and satisfy the normalization
condition $|p|^2+|q|^2=1$.
The dimensionless quantities which characterize mixing are
$x=2(m_2 - m_1)/(\Gamma_1+\Gamma_2)$ and 
$y=(\Gamma_2-\Gamma_1)/(\Gamma_1+\Gamma_2)$, where $m_{1,2}$ and 
$\Gamma_{1,2}$ are the masses and widths of the mass eigenstates. 
In the limit of $ \CP $ symmetry, $p$ and $q$ are equal.
To the extent that $CPV$ results only from $ p \neq q $,
and not from direct $CPV$ in the  $ D $ decay amplitudes themselves,
and in the limit $ | 1- | q/p |  | \,  \ll 1  $,
Wolfenstein's superweak constraint relates
the mixing and $CPV$ parameters~\cite{Wolfenstein:1964ks,Kagan:2009gb}:
\begin{equation}
\label{eqn:superweak}
\tan \varphi = \left( 1 - \left| { \frac{ q}{ p}  } \right| \right)
{ x \over y },
\end{equation}
where $ \varphi = \arg ( q/p ) $.
Allowing for both direct and indirect $CPV$, existing measurements 
give $ x = (0.37 \pm 0.16)\% $ and 
$ y = (0.66^{+0.07}_{-0.10})\% $~\cite{HFAG}.
These values are consistent 
with Standard Model (SM) expectations for long-distance
contributions~\cite{PhysRevD.69.114021,PhysRevD.33.179}.
No evidence for $CPV$ in mixing rates has been reported, and
SM expectations are $\leq 10^{-3}$~\cite{PhysRevD.65.054034,PhysRevD.69.114021,PhysRevD.33.179,Bianco:2003vb,Burdman:2003rs}.

We use  $\Dz\to K^\pm\pi^\mp$ decays to study mixing and $CPV$.
The decays $\Dz\to\Km\pip$ 
are called  ``right sign'' (RS) and
their decay rate is dominated by Cabibbo-favored (CF) amplitudes 
where no direct $CPV$ is expected in the SM or most of its extensions.
Decays of $\Dz\to\Kp\pim$ are called ``wrong sign'' (WS). Such decays
do not have such a simple description.
In the limit $(x, \, y) \ll 1 $, 
an approximation of the WS decay rates of 
the \Dz and \Dzb mesons is
\begin{eqnarray}
\label{eqn:d0mix}
\left|\bra{ K^+ \pi^-}H\dzkett\right|^2
& \approx & \frac{e^{-\Gamma t}}{2}\,| {\cal A}^{}_{ {f}}|^2
\biggl\{ {R}^{+}_D + \left|\frac{q}{p}\right|\,\sqrt{{R}^{+}_D}\,
\Bigl[ y\cos(\delta-\varphi) - x\sin(\delta-\varphi)\Bigr](\Gamma t) 
\nonumber \\
 & \ + \ & 
\left|\frac{q}{p}\right|^2
\frac{x^2+y^2}{4}\,(\Gamma t)^2 \biggr\} \label{eqn:d0mixing2}
\end{eqnarray}
and
\begin{eqnarray}
\label{eqn:d0barmix}
\left|\bra{ K^- \pi^+}H\dzbarkett\right|^2
& \approx & \frac{e^{-\Gamma t}}{2}\,|\overline {\cal A}^{}_{\bar f}|^2
\biggl\{ R^{-}_D + \left|\frac{p}{q}\right|\,\sqrt{R^{-}_D}\,
\Bigl[ y\cos(\delta+\varphi) - x\sin(\delta+\varphi)\Bigr](\Gamma t) 
\nonumber \\
 & \ + \  & 
\left|\frac{p}{q}\right|^2
\frac{x^2+y^2}{4}\,(\Gamma t)^2 \biggr\}  \label{eqn:d0mixing1} \,.
\end{eqnarray} 
In Eqs.~(\ref{eqn:d0mix})~and~(\ref{eqn:d0barmix}), $A_f$ denotes
the CF transition amplitude for $\Dz\to K^- \pi^+$ and 
$\overline{\mathcal{A}}_{\bar{f}}$ denotes the CF transition amplitude for $\Dzb\to K^+ \pi^- $.
The term $\Gamma$ is the average decay width of the two mass eigenstates.
Denoting the corresponding doubly Cabibbo-suppressed (DCS) amplitudes
$ {\cal A}_{\bar f} $ for $\Dz\to K^+\pi^-$ and $ \overline {\cal A}_{f} $ for $\Dzb\to K^-\pi^+$,
the ratios of DCS to CF amplitudes are defined to be
$ R_D^{+} =  | {\cal A}_{\bar f} / {\cal A}_f | ^2 $ and
$ R_D^{-} =  | \overline {\cal A}_f / \overline {\cal A}_{\bar f} | ^2 $.
The relative strong phase between the DCS and CF amplitudes
$ \overline{\cal A}_f $ and $ {\cal A}_f $ is denoted by $ \delta $.
We explicitly ignore direct $CPV$ in the phases of the CF and DCS amplitudes.
As the decay time $ t $ approaches zero, the WS rate is dominated 
by DCS amplitudes, where no direct $CPV$ is expected.
At longer decay times,
CF amplitudes associated with the corresponding antiparticle
produce oscillations; 
by themselves, they produce a 
pure mixing rate proportional to  $ (\Gamma t)^2 $,
and in combination with the DCS amplitudes
they produce an interference rate proportional to $ ( \Gamma t ) $.
Allowing for all possible types of $CPV$, the time-dependent ratio
of WS to RS decay rates, assuming $|x|\ll 1$ and $|y|\ll 1$, 
can be written as~\cite{Kagan:2009gb}
\begin{equation}
R(t)^\pm = R_D^\pm + \sqrt{R_D^\pm}y'^\pm\left( \frac{t}{\tau}\right) + \frac{(x'^{\pm})^2 + (y'^{\pm})^2}{4}\left( \frac{t}{\tau}\right)^2,
\label{eq:theeq}
\end{equation}
where the sign of the exponent in each term
denotes whether the decay is tagged at production as 
\Dz $(+)$ or as \Dzb $(-)$. The terms $x'$ and $y'$ are $x$ and $y$ 
rotated by the strong phase difference $\delta$, and $\tau = 1/\Gamma$.

The measured ratios of WS to RS decays differ from those of an
ideal experiment due to matter interactions, detector response and experimental misidentifications.
We use
the formal approach of Ref.~\cite{LHCb-PAPER-2013-053} to relate the signal
ratios of Eq.~(\ref{eq:theeq}) to a prediction of the experimentally observed ratios:
\begin{equation}
R(t)^\pm_{\text{pred}} = R(t)^\pm \left(1-\Delta_p^\pm \right)\left( \epsilon_r \right)^{\pm1} +p_{\text{other}},
\label{eq:eq2}
\end{equation}
where the term $\epsilon_r\equiv \epsilon(\Kp\pim)/\epsilon(\Km\pip)$ is the 
ratio of  \Kpm \pimp detection efficiencies. The efficiencies 
related to the $\pi_s^\pm$ and $\mu^\mp$ candidates explicitly cancel in this ratio.
The term $\Delta_p^\pm$ describes 
charge-specific peaking backgrounds produced by prompt charm
mistakenly included in the DT sample, assumed to be
zero after the ``same-sign background subtraction" described in Sec.~\ref{sec:selection}.
The term $p_{\text{other}}$ describes
peaking backgrounds that contribute differently to RS and WS decays. 
All three of these terms are considered to be potentially time dependent.

\section{Detector and Trigger}
\label{sec:Detector}

The \lhcb detector~\cite{Alves:2008zz,LHCb-DP-2014-002} is a single-arm forward
spectrometer covering the \mbox{pseudorapidity} range $2<\eta <5$,
and is designed for the study of particles containing \bquark or \cquark
quarks. The detector includes a high-precision tracking system
consisting of a silicon-strip vertex detector surrounding the $pp$
interaction region, a large-area silicon-strip detector located
upstream of a dipole magnet with a bending power of about
$4{\mathrm{\,Tm}}$, and three stations of silicon-strip detectors and straw
drift tubes placed downstream of the magnet.
The tracking system provides a measurement of momentum, \ptot, of charged particles with
a fractional uncertainty that varies from 0.5\% at 5\gevc to 1.0\% at 200\gevc.
The minimum distance of a track to a primary vertex (PV), the impact parameter (IP),
is measured with a resolution of $(15+29/\pt)\mum$,
where \pt is the component of the momentum transverse to the beam, in\,\gevc.
Different types of charged hadrons are distinguished using information
from two ring-imaging Cherenkov  (RICH) detectors.
Photons, electrons and hadrons are identified by a calorimeter system consisting of
scintillating-pad and preshower detectors, an electromagnetic
and a hadronic calorimeter. Muons are identified by a
system composed of alternating layers of iron and multiwire
proportional chambers.

The on-line candidate selection is performed by a trigger~\cite{LHCb-DP-2012-004} 
which consists of a hardware stage, based on information from the calorimeter and muon
systems, followed by a software stage.
At the hardware stage, candidates are required to have a muon with $\pt>$~1.64\gevc(1.76\gevc) 
in the 2011 (2012) data sets.
The software trigger, in which all charged particles
with $\pt>500\,(300)\mevc$ are reconstructed for 2011\,(2012) data, 
first requires a muon with $\pt>1.0$\gevc, and 
a large \chisqip with respect to any
PV, where \chisqip is defined as the
difference in vertex fit \chisq of a given PV reconstructed with and
without the muon. Following this selection, the muon and at least one other
final-state particle are required to be consistent with the topological signature
of the decay of a \bquark hadron~\cite{LHCb-DP-2012-004}. To mitigate detector-related asymmetries, the magnet polarity
is reversed periodically.

\section{Off-line selection}
\label{sec:selection}
In the off-line selection, candidates must have a muon with $p > 3 $\gevc,
$\pt > 1.2$\gevc and a track fit $\chisqndf<4$, where ndf is the number of degrees of freedom in the fit. 
Each of the \Dz decay products and muon candidates must
have \chisqip$>9$, consistent with originating from a secondary vertex. 
The slow pion candidate must have
$p > 2$\gevc and $\pt > 300$\mevc, 
and have no
associated hits in the muon stations. 
The combination of the $K$ and $\pi$ into
a \Dz candidate must form a vertex that is 
well separated from the PV and have 
a $\chi^2$ per degree of freedom less than 6.
The \Dz candidate must also have $\pt>1.4$\gevc and 
its reconstructed invariant mass must lie within 24\mevcc of its measured
mass~\cite{PDG2014}.
The
\Dstarp\mun invariant mass must lie in the range 3.1--5.1\gevcc.
Candidates must satisfy a 
vertex fit which 
constrains the kaon and pion
to come from the same vertex, and the muon, the slow pion and the \Dz
candidate to come from a common vertex with a good \chisqndf.
All final-state particles must pass stringent particle 
identification criteria from the RICH 
detectors, calorimeters and muon stations 
to improve the separation between
signal and backgrounds produced by misidentified final-state particles.
Candidates with reconstructed decay time $t/\tau<-0.5$ are vetoed, where $\tau$ is the measured
\Dz lifetime~\cite{PDG2014} and $t$ is calculated as $t=m_{\Dz} L/p$, where $L$ is the distance between
the \Dz production and decay vertices, $m_{\Dz}$ is the observed candidate \Dz mass, and $p$ is the \Dz momentum.
The decay-time resolution is roughly 120 fs for the DT sample.
Candidates which appear in both this data set and that of
the earlier prompt analysis~\cite{LHCb-PAPER-2013-053}
are vetoed. 

The same $ D^0 $ may appear in multiple candidate decay chains.
In about 0.5\% of cases,  a single $ \Dz \mun $ 
combination has multiple slow pion candidates whose laboratory
momentum vector directions lie within 0.6 mrad of each other.
In such cases, we randomly accept one of the candidates and
discard the others.
When two slow pion candidates associated with a single
$ \Dz\mun  $ candidate are not collinear,
the distributions of the $ D^0 \pi_s^+ $ masses are consistent with the
hypothesis that they
(typically) result from candidates with a real $ D^{*+} $ decay plus an 
additional pion nearby in phase space.
In such cases, we retain the multiple candidates; the fit
described below correctly determines the signal and background
rates as functions of $ m(D^0 \pi_s) $.

Real $ D^{*+} $ decays,
produced either promptly or as decay products of \bquark-hadron decays,
can be mistakenly associated with 
muons not truly originating from
$ \bquark $-hadron decays. 
In these cases, the production vertex of the $ D^0 $
may be wrongly determined.
We remove these from the $ D^0 \pi_s^+ $ distributions statistically
by subtracting the 
corresponding $ D^0 \pi_s^+ $ distributions of candidates where the $ D^{*+} $
and $ \mu $ candidates have the same charge, the so-called same-sign samples.
Signal candidates are referred to as the opposite-sign sample.
The  $ m(D^0 \pi_s^+ ) $ shapes to be subtracted are taken directly
from the same-sign candidates, while otherwise satisfying all DT selection criteria.
The absolute numbers of candidates
are determined, in each bin of the \Dz decay time, by
normalizing the same-sign rate to that of the opposite-sign
DT sample in the $m(\Dstarp\mun)$ range 5.6--6\gevcc, a region
well above the masses of the $B^0$ and $B_s^0$ mesons and dominated by
combinatorial backgrounds produced by false muon candidates.
The ratio of same-sign to DT candidates in the signal region is roughly 1\% and
the ratio in the normalization region is 71\%.
A systematic uncertainty on the same-sign background subtraction is determined
by setting the normalization factor to unity.

\section{Yield extraction and fit strategy}

\begin{figure}[htp]
\begin{center}
\includegraphics[width=0.48\textwidth]{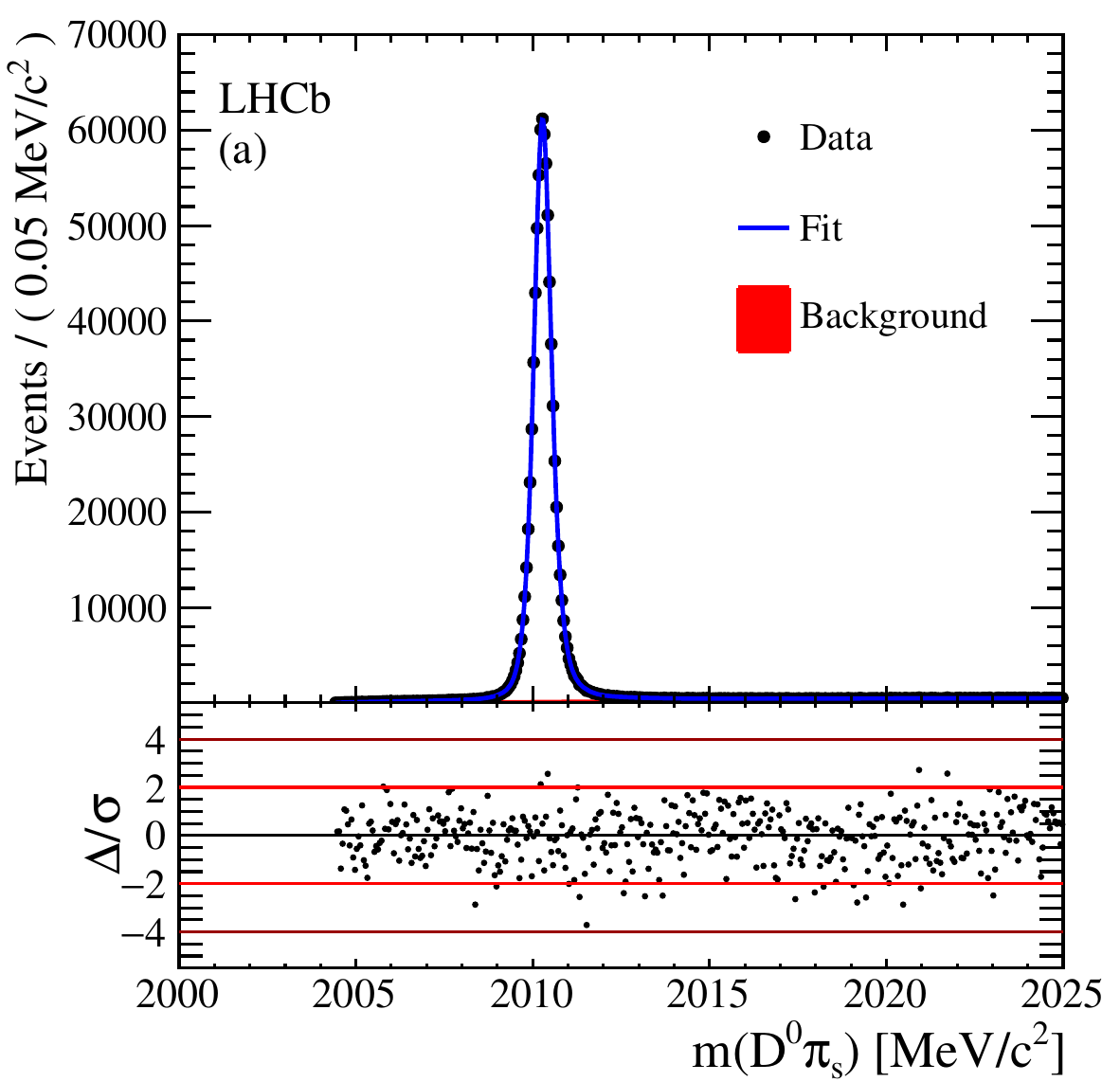}
\includegraphics[width=0.48\textwidth]{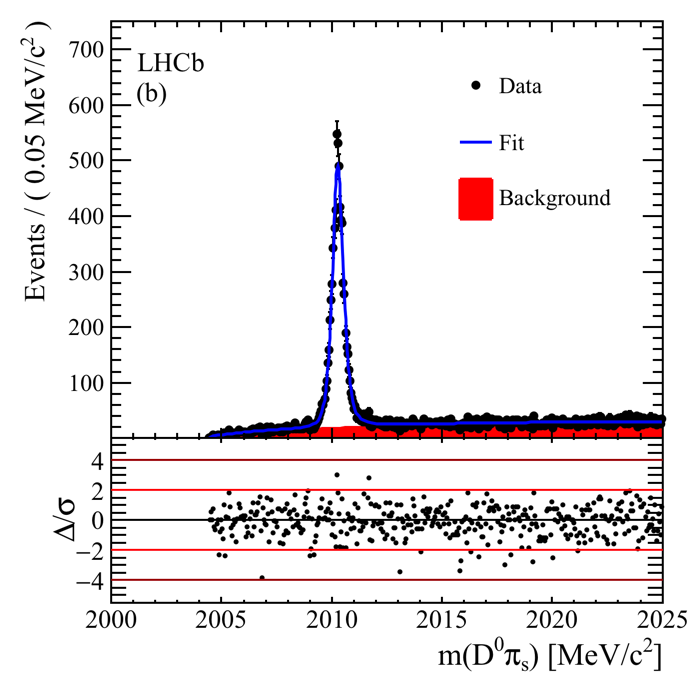}
\caption{The time-integrated $\Dz\pi^+_s$ invariant mass distributions,
after same-sign subtraction, for 
(a) RS decays and (b) WS decays. 
Fit projections are overlaid. Below each plot
are the normalized residual distributions.}
\label{fig:fits}
\end{center}
\end{figure}

Five bins of decay time are defined containing approximately
equal numbers of RS decays. We determine $ D^{*+} $
signal yields using binned maximum likelihood fits to the $\Dz\pi^+_s$ 
invariant mass distributions.
The signal probability density function (PDF) consists of a sum of three
Gaussian functions plus a Johnson $S_U$
distribution~\cite{johnsonSUfn} to model the asymmetric tails;
the background PDFs are parametrized using
empirical shapes of the form 
\begin{align}
{(m(\Dz\pi^+_s)/m_0-1)e^{c(m(\Dz\pi^+_s)/m_0-1)}}.
\end{align} 
The parameter $m_0$
represents the kinematic limit of the distribution and is fixed to the sum
of the measured mass of the pion and the \Dz~\cite{PDG2014}.
The shapes of the RS and WS $D^{* \pm}$ 
are assumed to be the same and to be independent of the decay time.
We first fit the time-integrated RS distribution
to determine signal shape parameters.
These are fixed for all subsequent fits.
The background parameters
vary independently in each fit. 
Systematic uncertainties related to this choice are assessed and discussed
in Sec~\ref{sec:systematics}.
Figures~\ref{fig:fits}(a)~and~\ref{fig:fits}(b) show the fits to the $\Dz \pi_s^+$ 
time-integrated invariant mass distributions for RS and WS
samples. 
They contain
$1.73\times 10^6$  and $6.68\times10^3$ 
$ D^{*+} $ decays, respectively.

The numbers of RS and WS signal candidates
in each decay time bin are determined from fits, from which
the observed WS to RS ratios are calculated.
The term $p_\text{other}$ is the ratio of
the number of peaking events in the $m(\Dz\pi_s^+)$
distribution from the $\Dz$ sidebands of the WS 
sample projected into the signal region
relative to the RS yield. We measure $p_\text{other}$ 
to be $(7.4\pm 1.8)\times 10^{-5}$. 
To measure the mixing and $CPV$ parameters,
the time dependence of these ratios is
fit by minimizing

\begin{equation}
\label{eqn:chis2Fit}
\chisq = \sum_{ i }\left[
\left(\frac{r_i^+-\widetilde{R}(t_i)^+}{\sigma_i^+}\right)^2 
+ \left(\frac{r_i^--\widetilde{R}(t_i)^-}{\sigma_i^-}\right)^2\right]
+\chisq_\epsilon +\chisq_{\text{peaking}} + \chisq_{\text{other}}.
\end{equation}

\noindent
Here, $r_i^\pm$ is the measured WS$^\pm$/RS$^\pm$ ratio for either the $\Dstarp(\Dz)$
or the $\Dstarm(\Dzb)$ sample with 
error $\sigma^\pm_i$ in a decay time bin $t_i$
and  $\widetilde{R}(t_i)$ is the value of
$R(t)^\pm_\text{pred}$ averaged over the bin. 
The fit accounts for uncertainties in the relative $ K^{\pm} \pi^{\mp} $ 
tracking and reconstruction efficiencies and rates of
peaking  backgrounds  using  Gaussian constraints
($ \chisq_\epsilon +\chisq_{\text{peaking}} + \chisq_{\text{other}} $).
The term $\chisq_{\text{other}}$ relating to the feedthrough
of the prompt sample into the DT sample is explicitly zero in the DT analysis,
but is needed for the simultaneous fit to the DT and prompt data sets.
The statistical uncertainties reported by the fit therefore include
the uncertainties associated with how precisely these factors
are determined.

Three fits are performed using this framework. 
First, we fit the data assuming \CP symmetry in the formalism of 
Eq.~(\ref{eq:theeq}) [{\em i.e.} $R^+ = R^- $, $(x'^{+})^2 = (x'^{-})^2$ 
and $ y'^+ =y'^- $]. 
Second, we fit the data requiring \CP symmetry in the CF and DCS 
amplitudes ({\em i.e.} $R^+ = R^-$),
but allow $CPV$ in the mixing parameters themselves [$ (x^{\prime \pm})^2$ and
$ y^{\prime \pm}$].
Finally, we fit the data allowing all the parameters to float freely.

\section{Relative efficiencies}

The relative efficiency $\epsilon_r $, used
in Eq.~(\ref{eq:eq2}), accounts for instrumental 
asymmetries in the $K^\mp\pi^\pm$ reconstruction efficiencies.
The largest source of these is the difference between 
the inelastic cross sections of $K^-$  and $ \pi^- $
mesons with matter, and those of their their antiparticles.
We measure $ \epsilon_r $, accounting for all detector
effects as well as cross-section differences in a similar manner to the prompt analysis~\cite{LHCb-PAPER-2013-053}.
The efficiency is
determined using the product of $D^+ \to K^-\pi^+\pi^+$ and 
$D^+ \to \KS (\to \pi^+\pi^-)\pi^+$ decay yields divided by the 
product of the corresponding charge-conjugate decay yields. 
The expected $CPV$ associated with differing $\Kz\to\KS$ and $\Kzb\to\KS$ rates and 
the differences in neutral kaon inelastic cross-sections with matter are accounted for~\cite{LHCb-PAPER-2014-013}.
Trigger and detection asymmetries
associated with the muon candidates
are calculated directly from data and included in the determination.
The $1\%$ asymmetry between \Dp and \Dm production 
rates~\cite{LHCb-PAPER-2012-026} cancels in this ratio, 
provided that the kinematic distributions are consistent across samples. 
To ensure this cancellation, we weight the $D^+ \to K^- \pi^+ \pi^+$ candidates 
such that the kaon $\pt$ and $\eta$ and pion $\pt$ distributions 
match those in the DT $K\pi$ sample. 
Similarly, $D^+ \to \KS \pi^+$ candidates are weighted by $D^+$ $\pt$ and $\eta$ and pion $\pt$ 
distributions to match those of the $\Dp\to\Km\pip\pip$.
The weighting is performed
using a gradient boosted decision tree implemented in
\textsc{scikit-learn}~\cite{Pedregosa:2012toh} accessed using the
\textsc{hep\_ml} framework~\cite{hepmlcite}.
We measure the $ K \pi $ detection asymmetry to be
$A(K\pi) = (\epsilon_r -1)/(\epsilon_r+1)  =  (0.90\pm0.18\pm0.10)\% $ for the
sample of this analysis,
and find it to be independent of decay time.

\begin{table}[thb]
\caption{Summary of systematic uncertainties for the DT analysis for each of the three fits described in the text.}
\begin{center}
\resizebox{\textwidth}{!}{
\begin{tabular}{lcccccc}
\hline
Source of systematic uncertainty &\multicolumn{6}{c}{Uncertainty on parameter} \\ \hline
& & & & & & \\
\multicolumn{7}{c}{No $CPV$}\\
& $R_D[10^{-3}]$ & $y'[10^{-3}]$ & $x'^2[10^{-4}]$& & &\\ \hline
 $\Dstarp\mup$ scaling        & 0.01 & 0.04 & 0.04& & &\\
 $A(K\pi)$ time dependence    & 0.01 & 0.07 & 0.04& & &\\
 RS fit model time variation & 0.00 & 0.01 & 0.03& & &\\ 
 No prompt veto              & 0.01 & 0.16 & 0.09& & &\\ \hline
 Total                       & 0.01 & 0.18 & 0.11& & &\\ 

& & & & & & \\
\multicolumn{7}{c}{No direct $CPV$}\\
 & $R_D[10^{-3}]$ & $y'^+[10^{-3}]$ & $\left(x'^{+}\right)^2[10^{-4}]$ & $y'^-[10^{-3}]$ & $\left(x'^{-}\right)^2[10^{-4}]$ &\\ \hline
 $\Dstarp\mup$ scaling        & 0.01 & 0.04 & 0.04 & 0.03 & 0.04 &\\
 $A(K\pi)$ time dependence    & 0.01 & 1.17 & 0.98 & 1.64 & 1.67 &\\
 RS fit model time variation & 0.00 & 0.02 & 0.03 & 0.01 & 0.03 &\\
 No prompt veto              & 0.01 & 0.11 & 0.00 & 0.19 & 0.19&\\ \hline
 Total                       & 0.01 & 1.17 & 0.98 & 1.66 & 1.68 &\\ 

& & & & & & \\
\multicolumn{7}{c}{All $CPV$ allowed}\\
& $R_D^+[10^{-3}]$ & $y'^+[10^{-3}]$ & $\left(x'^{+}\right)^2[10^{-4}]$& $R_D^-[10^{-3}]$ & $y'^-[10^{-3}]$ & $\left(x'^{-}\right)^2[10^{-4}]$\\ \hline
 $\Dstarp\mup$ scaling                  & 0.01 & 0.03 & 0.04 & 0.01 & 0.04 & 0.04\\ 
 $A(K\pi)$ time dependence    & 0.06 & 0.25 & 0.03 & 0.07 & 0.28 & 0.03\\ 
 RS fit model time variation & 0.00 & 0.01 & 0.01 & 0.00 & 0.04 & 0.05\\ 
 No prompt veto              & 0.01 & 0.09 & 0.01 & 0.01 & 0.21 & 0.19\\ 
 Potential fit biases           & 0.00 & 0.18 & 0.30 & 0.00 & 0.18 & 0.33\\ \hline
 Total                       & 0.06 & 0.32 & 0.31 & 0.07 & 0.40 & 0.38\\ \hline
\end{tabular}
}
\end{center}
\label{tab:systs}
\end{table}
\section{Systematic uncertainties}
\label{sec:systematics}
The systematic uncertainties of the DT analysis are summarized in Table~\ref{tab:systs}.
To avoid bias, offsets to each WS/RS ratio were randomly chosen to
blind both direct and indirect $CPV$, as well as the central values of the 
mixing parameters. 
Cross-checks of the blinded data were performed by splitting 
the data into disjoint subsamples according to criteria that might be sensitive to
systematic variations in detector response.
We considered 
two subsamples of magnet polarity, integrated  over the entire data taking
period, two subsamples for the year in which the data was recorded, 
four subsamples splitting according to magnet polarity and year of data acquisition, 
three subsamples each of $ K^\pm $ momentum,  $\mu^\pm $ transverse momentum
and  $ \pi^\pm $ transverse momentum. 
All observed variations in the fit results are consistent 
with being statistical fluctuations.

The ratio of RS \Dstarm to RS \Dstarp decays 
as a function of decay time 
is consistent with the hypothesis of decay-time independence 
with a $p$~value of $ 0.06 $.
We conservatively 
estimate a systematic uncertainty by modifying $ \epsilon_r $
to allow for a linear time dependence 
that gives the best description of the RS data.
As seen in Table \ref{tab:systs},
this has essentially no effect on the results of the mixing fit
where \CP symmetry is assumed to be exact.
It is the dominant systematic uncertainty in the fit 
requiring $ R_D^+ = R_D^- $, and it produces a systematic uncertainty much  
smaller than the statistical error in the fit that allows all forms of $CPV$.
Uncertainties are not symmetrized.

We determine systematic uncertainties related to
variations of the fit procedure by considering
alternative choices.
To determine uncertainties related to
subtraction of the same-sign background $ m (D^0 \pi_s )$ distributions from the opposite-sign ones,
we subtract the raw same-sign distributions rather than the
scaled distributions.
To determine uncertainties related to excluding
candidates considered in the prompt analysis~\cite{LHCb-PAPER-2013-053} from the DT
analysis,
we repeat the DT analysis including those candidates.
As an alternative to using a single signal shape at
all decay times, we determine signal shapes using the
RS signal in each decay time bin.
We evaluate potential biases in fitting procedure by generating and fitting
11,000 simulated DT samples with values of $ x $ and $ y $
spanning the 2$\sigma$ contour about the average values
reported by HFAG~\cite{HFAG}.
The biases we observe are nonzero, and appear to
be independent of the generated values. 
We assign
a systematic uncertainty equal to the full observed bias.
Table~\ref{tab:systs} summarizes the results of these
studies.
All systematic uncertainties are propagated to 
the final results using their full covariance matrices.


\begin{figure}[hbt]
\begin{center}
\includegraphics[width=0.5\textwidth]{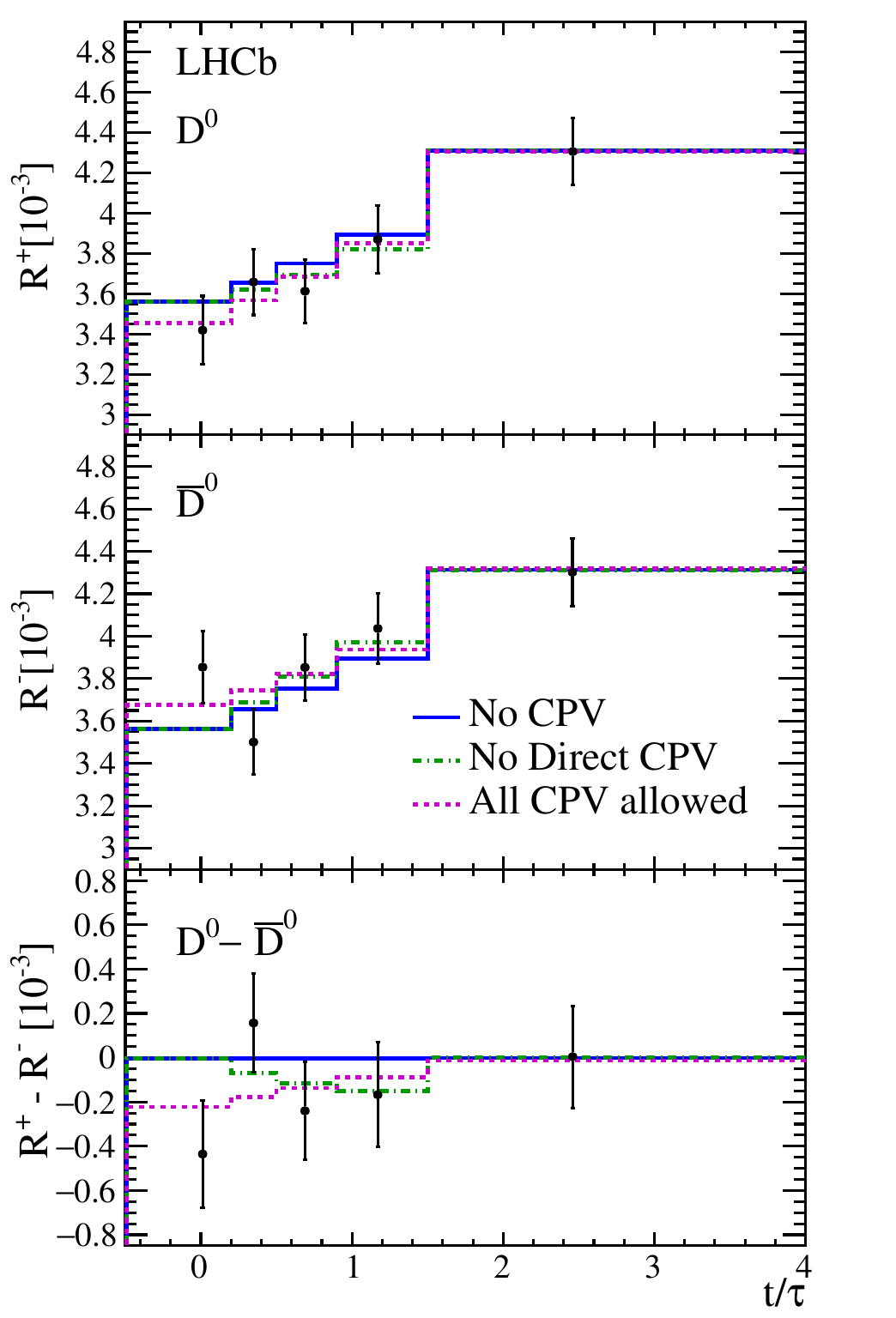}
\caption{Efficiency corrected and SS background subtracted
ratios of WS/RS decays 
and fit projections for the DT sample. 
The top plot shows the  \Dz $(R^+(t))$ sample.
The middle plot shows the \Dzb $(R^-(t))$ sample.
The bottom plot shows the difference between the top
and middle plots.
In all cases, the error bars superposed on the data
points are those from the $ \chi^2 $ minimization fits
with no accounting for additional systematic
uncertainties.
The projections shown are for fits assuming \CP symmetry (solid blue line),
allowing no direct $CPV$ (dashed-dotted green line), and allowing all forms of $CPV$ (dashed magenta line). 
Bins are centered at the average value of $t/\tau$
of the bin.
}
\label{fig:fitproj}
\end{center}
\end{figure}

\section{Results}

The efficiency-corrected and same-sign (SS) background subtracted
WS/RS ratios of the DT data and the three fits described earlier are shown in Fig.~\ref{fig:fitproj}.
The fits are shown in a binned projection.
The top two plots show the WS/RS ratio as a function of decay time
for the candidates tagged at production as \Dz [$ R^+ (t) $]
and as \Dzb [$ R^- (t) $].
Both sets of points appear to lie on straight
lines that intersect the vertical axis near
$ 3.5 \times 10^{-3} $ at $ t / \tau = 0 $ and rise approximately
linearly to $ 4.3 \times 10^{-3} $ near $ t / \tau = 2.5 $.
The difference between the two ratios is shown in the bottom
plot.  
The fit values for the parameters and their uncertainties,
are collected in Table~\ref{tab:theDTres}.
The data are clearly consistent with the hypothesis of \CP symmetry, \ie that the
two samples share exactly the same mixing parameters.
If direct CPV is assumed to be zero ($R^+=R^-$ at $t/\tau=0$), 
as expected if tree-level
amplitudes dominate the CF and DCS amplitudes, the difference in 
mixing rates (the slope) is observed to be very small.
For this data set, the statistical uncertainties are all
much greater than the corresponding systematic uncertainties, which include
the uncertainties from $\epsilon_r$ and peaking backgrounds. Correlation matrices
between the fitted parameters are included in Appendix~\ref{app:dt_cov_mat_dt}, Tables~\ref{tab:final_cov_mat_mix}--\ref{tab:final_cov_mat_allcpv}.

{\renewcommand{\arraystretch}{1.1}
\begin{table}[htbp]
\caption{Fitted parameters of the DT sample. The first uncertainties include the statistical uncertainty, as well as
the peaking backgrounds and the $K\pi$ detection efficiency, and the second are systematic.}
\begin{center}
\begin{tabular}{l
  S[table-format=1.2]@{\,\( \pm \)\,}
  S[table-format=1.2]@{\,\( \pm \)\,}
  S[table-format=1.2]
  }
\hline
Parameter & \multicolumn{3}{c}{Value}\\ 
\hline 
\multicolumn{4}{c}{No CPV}\\\hline
$R_D [10^{-3}]$ & 3.48 & 0.10 & 0.01 \\ 
$x'^2 [10^{-4}]$ &  0.28 & 3.10 & 0.11 \\ 
$y' [10^{-3}]$ & 4.60 & 3.70 & 0.18 \\ 
\chisqndf &  \multicolumn{3}{c}{6.3/7}\\ 
\\\hline
\multicolumn{4}{c}{No direct CPV}\\\hline
$R_D [10^{-3}]$ & 3.48 & 0.10 & 0.01 \\ 
$\left(x'^{+}\right)^2 [10^{-4}]$ &  1.94 & 3.67 & 0.98 \\ 
$y'^{+} [10^{-3}]$ & 2.79 & 4.27 & 1.17 \\ 
$\left(x'^{-}\right)^2 [10^{-4}]$ &  -1.53 & 4.04 & 1.68 \\ 
$y'^{-} [10^{-3}]$ &  6.51 & 4.38 & 1.66\\ 
\chisqndf &  \multicolumn{3}{c}{5.6/5}\\ 
\\\hline
\multicolumn{4}{c}{All CPV allowed}\\\hline
$R_D^+ [10^{-3}]$ & 3.38 & 0.15 & 0.06 \\ 
$\left(x'^{+}\right)^2 [10^{-4}]$ &  -0.19 & 4.46 & 0.31\\
$y'^{+} [10^{-3}]$ & 5.81 & 5.25 &0.32\\ 
$R_D^- [10^{-3}]$ & 3.60 & 0.15 & 0.07 \\ 
$\left(x'^{-}\right)^2 [10^{-4}]$ &  0.79 & 4.31 & 0.38 \\ 
$y'^{-} [10^{-3}]$ & 3.32 & 5.21 & 0.40 \\ 
\chisqndf &  \multicolumn{3}{c}{4.5/4}\\  \hline 
\end{tabular}
\end{center}
\label{tab:theDTres}
\end{table}%
}

\begin{figure}[htb]
\begin{center}
\includegraphics[width=0.5\textwidth]{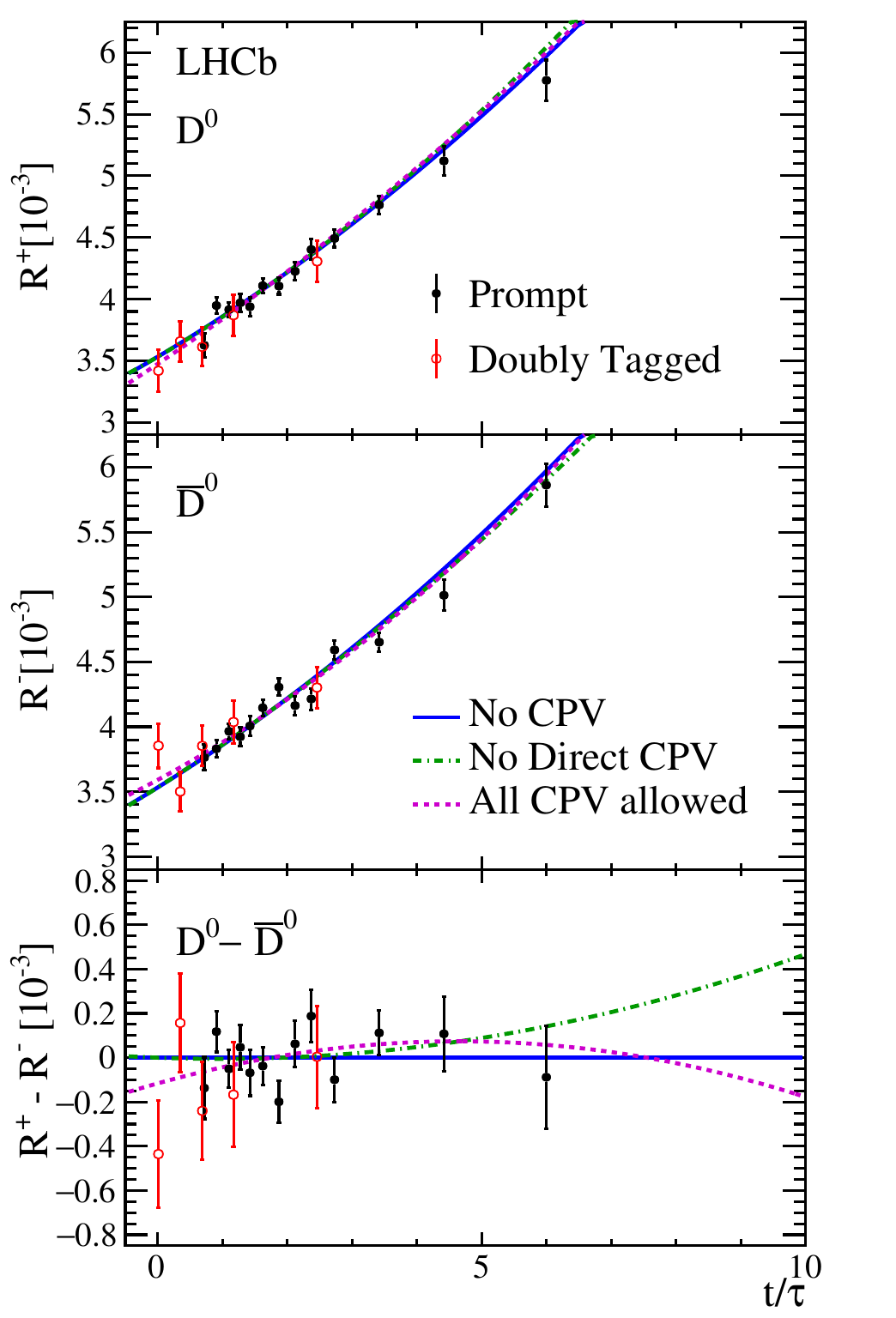}
\caption{
Efficiency-corrected data
and fit projections for the DT (red open circles) and prompt (black filled circles) samples.
The top plot shows the \Dz $(R^+(t))$ samples.
The middle plot shows the \Dzb $(R^-(t))$ samples.
The bottom plot shows the difference between the top
and middle plots. 
In all cases, the error bars superposed on the data
points are those from the $ \chi^2 $ minimization fits
without accounting for additional systematic
uncertainties.
The projections shown are for fits assuming \CP symmetry (solid blue line),
allowing no direct $CPV$ (dashed-dotted green line), and allowing all forms of $CPV$ (dashed magenta line). Bins are centered
at the average $t/\tau$ of the bin.}
\label{fig:totProj}
\end{center}
\end{figure}

The data of the prompt analysis~\cite{LHCb-PAPER-2013-053},
those of the DT analysis and the
results of fitting the two (disjoint)
samples simultaneously are shown in Fig.~\ref{fig:totProj}.
The combined sets of
data points in the top and middle plots lie on slightly
curved lines 
that intersect the vertical axis near
$ 3.4 \times 10^{-3} $ at $ t / \tau = 0 $ and rise to approximately
$ 5.9 \times 10^{-3} $ just above $ t / \tau = 6.0 $.
The samples are consistent with
\CP symmetry.
The results of the simultaneous fit are reported
in Table~\ref{tab:allRes}.
The corresponding results from the prompt analysis~\cite{LHCb-PAPER-2013-053}
are also reported in Table~\ref{tab:allRes} for comparison.
In Table~\ref{tab:allRes},
the statistical and systematic uncertainties have been added in
quadrature to allow direct comparison of the two sets of results.
As all the systematic uncertainties for the prompt
analysis were evaluated using $ \chi^2 $ constraints
as in Eq.~(\ref{eqn:chis2Fit}),
we determine systematic uncertainties for the simultaneous fits
by repeating the fit variations  
as for the DT fit. These systematics are reported in Table~\ref{tab:comb_systs}.
In general, the uncertainties from the combined fits are 10\%--20\%
lower than those from the previous measurement~\cite{LHCb-PAPER-2013-053}.
The decrease in the uncertainty comes from the improved precision that the DT sample
provides at low \Dz decay time.
The corresponding correlation matrices are given in Appendix~\ref{app:cov_mat_dt_prompt}, Tables~\ref{tab:cov_mat_dt_prompt_mix}--\ref{tab:cov_mat_dt_prompt_allcpv}.

The combined fit of the DT and prompt sample is consistent with
\CP symmetry.
The WS \Dz and \Dzb rates at $t/\tau=0$ are equal within experimental uncertainties,
indicating no direct \CP violation. Similarly, the mixing rates are consistent
within experimental uncertainties, as seen in the bottom plot of Fig.~\ref{fig:totProj}.
In the combined fit of this analysis, assuming no direct \CP violation,
the difference between the projected WS/RS rates at $t/\tau = 6.0$ is only $0.15 \times 10^{-3}$ (see the dashed-dotted
line in the bottom plot of Fig.~\ref{fig:totProj}), where the WS/RS rates themselves have increased by
about $2.5 \times 10^{-3}$ (see the top and middle plots).

The determination of the $CPV$ parameters $|q/p|$ and $\varphi$ from the difference in rates of WS \Dz and \Dzb 
requires the use of independent measurements,
as these variables appear in the WS/RS ratios only in combination with 
the strong phase difference $\delta$ and with $x$ and $y$, as seen in Eqs.~(\ref{eqn:d0mix})~and~(\ref{eqn:d0barmix}). 
When the results are combined with independent measurements, as done by the Heavy Flavor Averaging Group~\cite{HFAG},
the precision of the constraints on $ |q/p| - 1 $ approximately scale with
the precision of the difference in WS/RS ratios at high decay
time divided by the average increase. 
Utilizing theoretical constraints such as Eq.~(\ref{eqn:superweak}), in addition to the experimental data,
the precision on $|q/p|$ improves by about a factor of 4~\cite{HFAG}.

\section{Summary}

In summary, the analysis of mixing and $CPV$ 
parameters using the DT $ \Dz \to K^\mp \pi^\pm $
samples provides results consistent with those of our earlier prompt
analysis.
Simultaneously fitting the disjoint data sets of the two analyses
improves the precision of the measured parameters by 10\%--20\%, even 
though the DT analysis is based on almost 40 times fewer candidates than the 
prompt analysis.
In part, this results from much cleaner signals in the DT analysis,
and, in part, it results from the complementary
higher acceptance of the DT trigger at low $ D $ decay times.
The current results supersede
those of our earlier publication~\cite{LHCb-PAPER-2013-053}.

{\renewcommand{\arraystretch}{1.1}
\begin{table}[htbp]
\caption{Simultaneous fit result of the DT and prompt samples. The prompt-only results from~\cite{LHCb-PAPER-2013-053} are
shown on the right for comparison. Statistical and systematic errors have been added in quadrature.}
\begin{center}
\begin{tabular}{lcc}
\hline
Parameter & DT + Prompt & Prompt-only\\ \hline
\multicolumn{3}{c}{No $CPV$}\\\hline
$R_D [10^{-3}]$ & $ 3.533 \pm 0.054$  & $ 3.568 \pm 0.067$\\ 
$x'^2 [10^{-4}]$ & $ 0.36 \pm 0.43$ & $0.55 \pm 0.49$\\ 
$y' [10^{-3}]$ & $ 5.23 \pm 0.84$ & $ 4.8 \pm 0.9$\\ 
\chisqndf & 96.6/111& 86.4/101\\ 
\\\hline
\multicolumn{3}{c}{No direct $CPV$}\\\hline
$R_D [10^{-3}]$ & $ 3.533 \pm 0.054$ & $ 3.568\pm  0.067$\\ 
$\left(x'^{+}\right)^2 [10^{-4}]$ & $ 0.49 \pm 0.50$ &  $0.64 \pm 0.56$\\ 
$y'^{+} [10^{-3}]$ & $ 5.14 \pm 0.91$ & $4.8\pm 1.1$\\ 
$\left(x'^{-}\right)^2 [10^{-4}]$ & $ 0.24 \pm 0.50$ & $0.46\pm 0.55$\\
$y'^{-} [10^{-3}]$ & $ 5.32 \pm 0.91$ & $4.8 \pm 1.1$\\
\chisqndf & 96.1/109& 86.0/99\\ 
\\\hline
\multicolumn{3}{c}{All $CPV$ allowed}\\\hline
$R_D^+ [10^{-3}]$ & $ 3.474 \pm 0.081$ & $3.545 \pm 0.095$ \\ 
$\left(x'^{+}\right)^2 [10^{-4}]$ & $ 0.11 \pm 0.65$ & $ 0.49 \pm 0.70$\\ 
$y'^{+} [10^{-3}]$ & $ 5.97 \pm 1.25$ &$5.1\pm1.4$ \\ 
$R_D^- [10^{-3}]$ & $ 3.591 \pm 0.081$ &  $3.591 \pm 0.090$\\ 
$\left(x'^{-}\right)^2 [10^{-4}]$ & $ 0.61 \pm 0.61$ & $0.60\pm 0.68$\\ 
$y'^{-} [10^{-3}]$ & $ 4.50 \pm 1.21$ & $4.5\pm  1.4$\\ 
\chisqndf & 95.0/108 & 85.9/98\\ \hline
\end{tabular}
\end{center}
\label{tab:allRes}
\end{table}%
}

\begin{table}[htp]
\caption{Systematic uncertainties for the simultaneous fits of the DT and prompt datasets.}
\begin{center}
\resizebox{\textwidth}{!}{
\begin{tabular}{lcccccc}
\hline
Systematic uncertainty & \multicolumn{6}{c}{Uncertainty on parameter} \\ \hline
& & & & & & \\
\multicolumn{7}{c}{No $CPV$} \\ 
 & $R_D[10^{-3}]$ & $y'[10^{-3}]$ & $x'^2[10^{-4}]$ & & &\\ \hline
 $\Dstarp\mup$ scaling                  & 0.00 & 0.05 & 0.02& & &\\ 
 $A(K\pi)$ time dependence    & 0.00 & 0.02 & 0.01& & &\\ 
 RS fit model time variation & 0.00 & 0.00 & 0.00& & &\\ 
 No prompt veto              & 0.00 & 0.04 & 0.02& & &\\ \hline 
 Total                       & 0.01 & 0.07 & 0.03& & &\\ 

& & & & & & \\
 \multicolumn{7}{c}{No direct $CPV$} \\ 
& $R_D[10^{-3}]$ & $y'^+[10^{-3}]$ & $\left(x'^{+}\right)^2[10^{-4}]$ & $y'^-[10^{-3}]$ & $\left(x'^{-}\right)^2[10^{-4}]$&\\ \hline
 $\Dstarp\mup$ scaling                  & 0.00 & 0.05 & 0.02 & 0.05 & 0.02&\\ 
 $A(K\pi)$ time dependence    & 0.00 & 0.02 & 0.02 & 0.06 & 0.03&\\ 
 RS fit model time variation & 0.00 & 0.00 & 0.00 & 0.00 & 0.00&\\ 
 No prompt veto              & 0.00 & 0.04 & 0.02 & 0.04 & 0.02&\\ \hline
 Total                       & 0.01 & 0.07 & 0.03 & 0.09 & 0.05&\\ 

& & & & & & \\
\multicolumn{7}{c}{All $CPV$ allowed} \\ 
& $R_D^+[10^{-3}]$ & $y'^+[10^{-3}]$ & $\left(x'^{+}\right)^2[10^{-4}]$& $R_D^-[10^{-3}]$ & $y'^-[10^{-3}]$ & $\left(x'^{-}\right)^2[10^{-4}]$\\ \hline
 $\Dstarp\mup$ scaling                  & 0.00 & 0.06 & 0.02 & 0.00 & 0.05 & 0.02\\ 
 $A(K\pi)$ time dependence    & 0.03 & 0.33 & 0.15 & 0.03 & 0.31 & 0.13\\ 
 RS fit model time variation & 0.00 & 0.00 & 0.00 & 0.00 & 0.01 & 0.00\\ 
 No prompt veto              & 0.00 & 0.05 & 0.02 & 0.00 & 0.03 & 0.01\\ \hline
 Total                       & 0.03 & 0.34 & 0.15 & 0.03 & 0.31 & 0.13\\ \hline
\end{tabular}
}
\end{center}
\label{tab:comb_systs}
\end{table}%

\clearpage
\section*{Acknowledgements}

\noindent We express our gratitude to our colleagues in the CERN
accelerator departments for the excellent performance of the LHC. We
thank the technical and administrative staff at the LHCb
institutes. We acknowledge support from CERN and from the national
agencies: CAPES, CNPq, FAPERJ and FINEP (Brazil); NSFC (China);
CNRS/IN2P3 (France); BMBF, DFG and MPG (Germany); INFN (Italy);
FOM and NWO (Netherlands); MNiSW and NCN (Poland); MEN/IFA (Romania);
MinES and FASO (Russia); MinECo (Spain); SNSF and SER (Switzerland);
NASU (Ukraine); STFC (United Kingdom); and NSF (USA).
We acknowledge the computing resources that are provided by CERN, IN2P3 (France), KIT and DESY (Germany), INFN (Italy), SURF (Netherlands), PIC (Spain), GridPP (United Kingdom), RRCKI and Yandex LLC (Russia), CSCS (Switzerland), IFIN-HH (Romania), CBPF (Brazil), PL-GRID (Poland) and OSC (USA). We are indebted to the communities behind the multiple open
source software packages on which we depend.
Individual groups or members have received support from AvH Foundation (Germany),
EPLANET, Marie Sk\l{}odowska-Curie Actions and ERC (European Union),
Conseil G\'{e}n\'{e}ral de Haute-Savoie, Labex ENIGMASS and OCEVU,
R\'{e}gion Auvergne (France), RFBR and Yandex LLC (Russia), GVA, XuntaGal and GENCAT (Spain), Herchel Smith Fund, The Royal Society, Royal Commission for the Exhibition of 1851 and the Leverhulme Trust (United Kingdom).

\clearpage

\clearpage

{\noindent\normalfont\bfseries\Large Appendices}

\appendix

\section{Correlation matrices of the DT Fit}
\label{app:dt_cov_mat_dt}

\begin{table}[!htp]
\caption{Correlation matrix for the no $CPV$ fit to the DT data.}
\begin{center}
  \begin{tabular}{ c|ccc }
       & $R_D$ & $y'$                         & $x'^2$ \\ \hline
$R_D$  & 1     & $-0.786$                     & $\phantom{-}0.671          $ \\
$y'$   &       & $\phantom{-}1\phantom{.000}$ & $          -0.964          $ \\
$x'^2$ &       &                              & $\phantom{-}1\phantom{.000}$ \\
\end{tabular}
\end{center}
\label{tab:final_cov_mat_mix}
\end{table}%

\begin{table}[!htp]
\caption{Correlation matrix for the no direct $CPV$ fit to the DT data.}
\begin{center}
\begin{tabular}{ c|ccccc }
                        & $R_D$ & $y'^+$                       & $\left(x'^{+}\right)^2$      & $y'^-$                       & $\left(x'^{-}\right)^2$ \\ \hline
$R_D$                   &  1    & $-0.655$                     & $\phantom{-}0.523$           & $          -0.616          $ & $\phantom{-}0.494$ \\
$y'^+$                  &       & $\phantom{-}1\phantom{.000}$ & $          -0.892$           & $\phantom{-}0.398          $ & $          -0.319 $ \\
$\left(x'^{+}\right)^2$ &       &                              & $\phantom{-}1\phantom{.000}$ & $          -0.318          $ & $\phantom{-}0.255 $ \\
$y'^-$                  &       &                              &                              & $\phantom{-}1\phantom{.000}$ & $          -0.834 $ \\
$\left(x'^{-}\right)^2$ &       &                              &                              &                              & $\phantom{-}1\phantom{.000}$ \\
\end{tabular}
\end{center}
\label{tab:final_cov_mat_nodcpv}
\end{table}%

\begin{table}[!htp]
\caption{Correlation matrix for the all $CPV$ allowed fit to the DT data.}
\begin{center}
\begin{tabular}{ c|cccccc }
                         & $R_D^+$  & $y'^+$                       & $\left(x'^{+}\right)^2$      & $R_D^-$                      & $y'^-$                       & $\left(x'^{-}\right)^2$ \\
\hline
$R_D^+$                  &       1  & $-0.732$                     & $\phantom{-}0.625$           & $-0.008$                     & $\phantom{-}0.000$           & $\phantom{-}0.000          $ \\
$y'^+$                   &          & $\phantom{-}1\phantom{.000}$ & $          -0.963$           & $\phantom{-}0.000$           & $\phantom{-}0.000$           & $\phantom{-}0.000          $ \\
$\left(x'^{+}\right)^2$  &          &                              & $\phantom{-}1\phantom{.000}$ & $\phantom{-}0.000$           & $\phantom{-}0.000$           & $\phantom{-}0.000          $ \\
$R_D^-$                  &          &                              &                              & $\phantom{-}1\phantom{.000}$ & $          -0.707$           & $\phantom{-}0.602          $ \\
$y'^-$                   &          &                              &                              &                              & $\phantom{-}1\phantom{.000}$ & $          -0.958          $ \\
$\left(x'^{-}\right)^2$  &          &                              &                              &                              &                              & $\phantom{-}1\phantom{.000}$ \\
\end{tabular}
\end{center}
\label{tab:final_cov_mat_allcpv}
\end{table}%

\clearpage
\section{Correlation matrices of the DT + Prompt Fit}
\label{app:cov_mat_dt_prompt}

Please note that Ref.~\cite{LHCb-PAPER-2013-053} has been superseded by Ref.~\cite{LHCb-PAPER-2024-008}.
Therefore, the combination that Tables~8 to 10 refer to should no longer be used.

\begin{table}[htbp]
\caption{Correlation matrix for the no $CPV$ simultaneous fit to the prompt + DT data sets.}
\begin{center}
\begin{tabular}{ c|ccc }
   & $R_D$  & $y'$  & $x'^2$ \\\hline
  $R_D$  & 1  & $ -0.932 $  &$ \phantom{-}0.826 $ \\
  $y'$  &  & 1  & $ -0.959 $ \\
  $x'^2$  &  &  & 1 \\
\end{tabular}
\end{center}
\label{tab:cov_mat_dt_prompt_mix}
\end{table}%

\begin{table}[htbp]
\caption{Correlation matrix for the no direct $CPV$ simultaneous fit to the prompt + DT data sets.}
\begin{center}
\begin{tabular}{ c|ccccc }
   & $R_D$  & $y'^+$  & $\left(x'^{+}\right)^2$  & $y'^-$  & $\left(x'^{-}\right)^2$\\  \hline
  $R_D$  & 1  & $ -0.854 $  &$ \phantom{-}0.686 $  & $ -0.751 $  &$ \phantom{-}0.586 $ \\
  $y'^+$  &  & 1  & $ -0.925 $  &$ \phantom{-}0.631 $  & $ -0.501 $ \\
  $\left(x'^{+}\right)^2$  &  &  & 1  & $ -0.563 $  &$ \phantom{-}0.458 $ \\
  $y'^-$  &  &  &  & 1  & $ -0.937 $ \\
  $\left(x'^{-}\right)^2$  &  &  &  &  & 1 \\
\end{tabular}
\end{center}
\label{tab:cov_mat_dt_prompt_nodcpv}
\end{table}%

\begin{table}[htbp]
\caption{Correlation matrix for the all $CPV$ allowed simultaneous fit to the prompt + DT data sets.}
\begin{center}
\begin{tabular}{ c|cccccc }
   & $R_D^+$  & $y'^+$  & $\left(x'^{+}\right)^2$  & $R_D^-$  & $y'^-$  & $\left(x'^{-}\right)^2$\\ \hline
  $R_D^+$  & 1  & $ -0.920 $  &$ \phantom{-}0.823 $  & $ -0.007 $  & $ -0.010 $  &$ \phantom{-}0.008 $ \\
  $y'^+$  &  & 1  & $ -0.962 $  & $ -0.011 $  &$ \phantom{-}0.000 $  & $ -0.002 $ \\
  $\left(x'^{+}\right)^2$  &  &  & 1  &$ \phantom{-}0.009 $  & $ -0.002 $  &$ \phantom{-}0.004 $ \\
  $R_D^-$  &  &  &  & 1  & $ -0.918 $  &$ \phantom{-}0.812 $ \\
  $y'^-$  &  &  &  &  & 1  & $ -0.956 $ \\
  $\left(x'^{-}\right)^2$  &  &  &  &  &  & 1 \\
\end{tabular}
\end{center}
\label{tab:cov_mat_dt_prompt_allcpv}
\end{table}%

\clearpage

\clearpage
\addcontentsline{toc}{section}{References}
\setboolean{inbibliography}{true}
\ifx\mcitethebibliography\mciteundefinedmacro
\PackageError{LHCb.bst}{mciteplus.sty has not been loaded}
{This bibstyle requires the use of the mciteplus package.}\fi
\providecommand{\href}[2]{#2}

\newpage



\newpage
\centerline{\large\bf LHCb collaboration}
\begin{flushleft}
\small
R.~Aaij$^{40}$,
B.~Adeva$^{39}$,
M.~Adinolfi$^{48}$,
Z.~Ajaltouni$^{5}$,
S.~Akar$^{6}$,
J.~Albrecht$^{10}$,
F.~Alessio$^{40}$,
M.~Alexander$^{53}$,
S.~Ali$^{43}$,
G.~Alkhazov$^{31}$
% ...

{\footnotesize \it
% ...
$ ^{5}$Clermont Universit{\'e}, Universit{\'e} Blaise Pascal, CNRS/IN2P3, LPC, Clermont-Ferrand, France\\
$ ^{6}$CPPM, Aix-Marseille Universit{\'e}, CNRS/IN2P3, Marseille, France\\
% ...
$ ^{10}$Fakult{\"a}t Physik, Technische Universit{\"a}t Dortmund, Dortmund, Germany\\
% ...
$ ^{31}$Petersburg Nuclear Physics Institute (PNPI), Gatchina, Russia\\
% ...
$ ^{39}$Universidad de Santiago de Compostela, Santiago de Compostela, Spain\\
$ ^{40}$European Organization for Nuclear Research (CERN), Geneva, Switzerland\\
% ...
$ ^{43}$Nikhef National Institute for Subatomic Physics, Amsterdam, The Netherlands\\
% ...
$ ^{48}$H.H. Wills Physics Laboratory, University of Bristol, Bristol, United Kingdom\\
% ...
$ ^{53}$School of Physics and Astronomy, University of Glasgow, Glasgow, United Kingdom\\
% ...
}
\end{flushleft}

\newpage
\pagenumbering{gobble}

\end{document}


\pagestyle{plain} 
\setcounter{page}{1}
\pagenumbering{arabic}

\section{Supplementary material for LHCb-PAPER-2016-033}
\label{sec:Supplementary-App}

This appendix contains supplementary material that will posted
on the public CDS record but will not appear in the paper.


\begin{figure}[htbp]
\begin{center}
\includegraphics[width=0.7\textwidth]{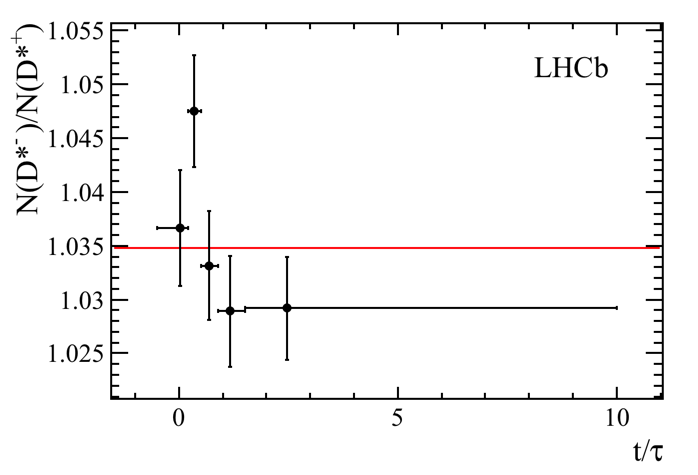}
\caption{Ratio of RS \Dstarm over \Dstarp yields as a function of $t/\tau$ of the \Dz. The red line is the fit to a constant. The result
of the fit gives a \chisqndf = 8.842/4, corresponding to a $p$-value of 6\%. The central value of the fit is $1.035\pm0.002$.}
\end{center}
\end{figure}

\begin{figure}[htbp]
\begin{center}
\resizebox{\textwidth}{!}{
\includegraphics[width=\textwidth]{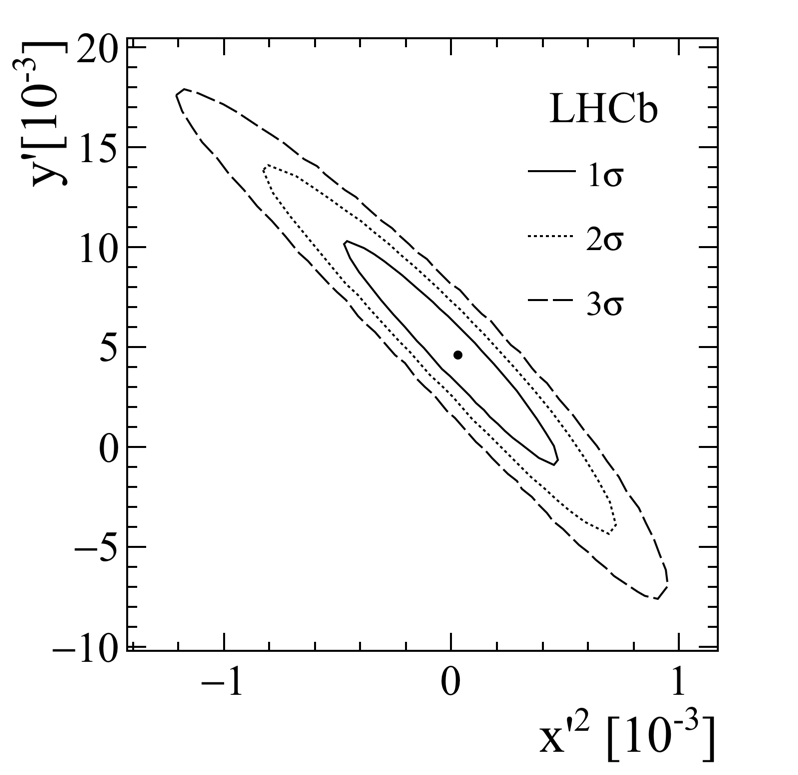}
\includegraphics[width=\textwidth]{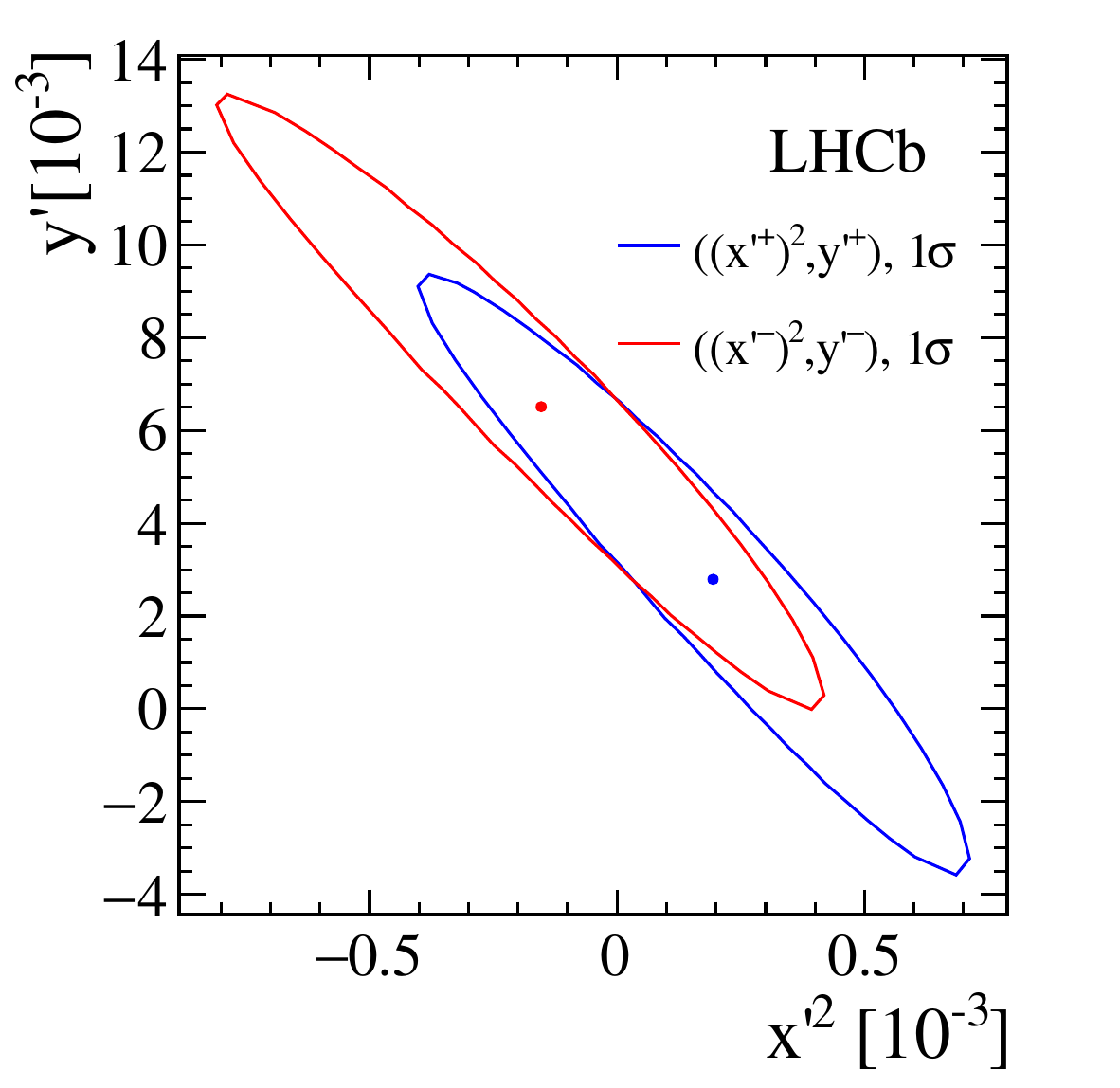}
\includegraphics[width=\textwidth]{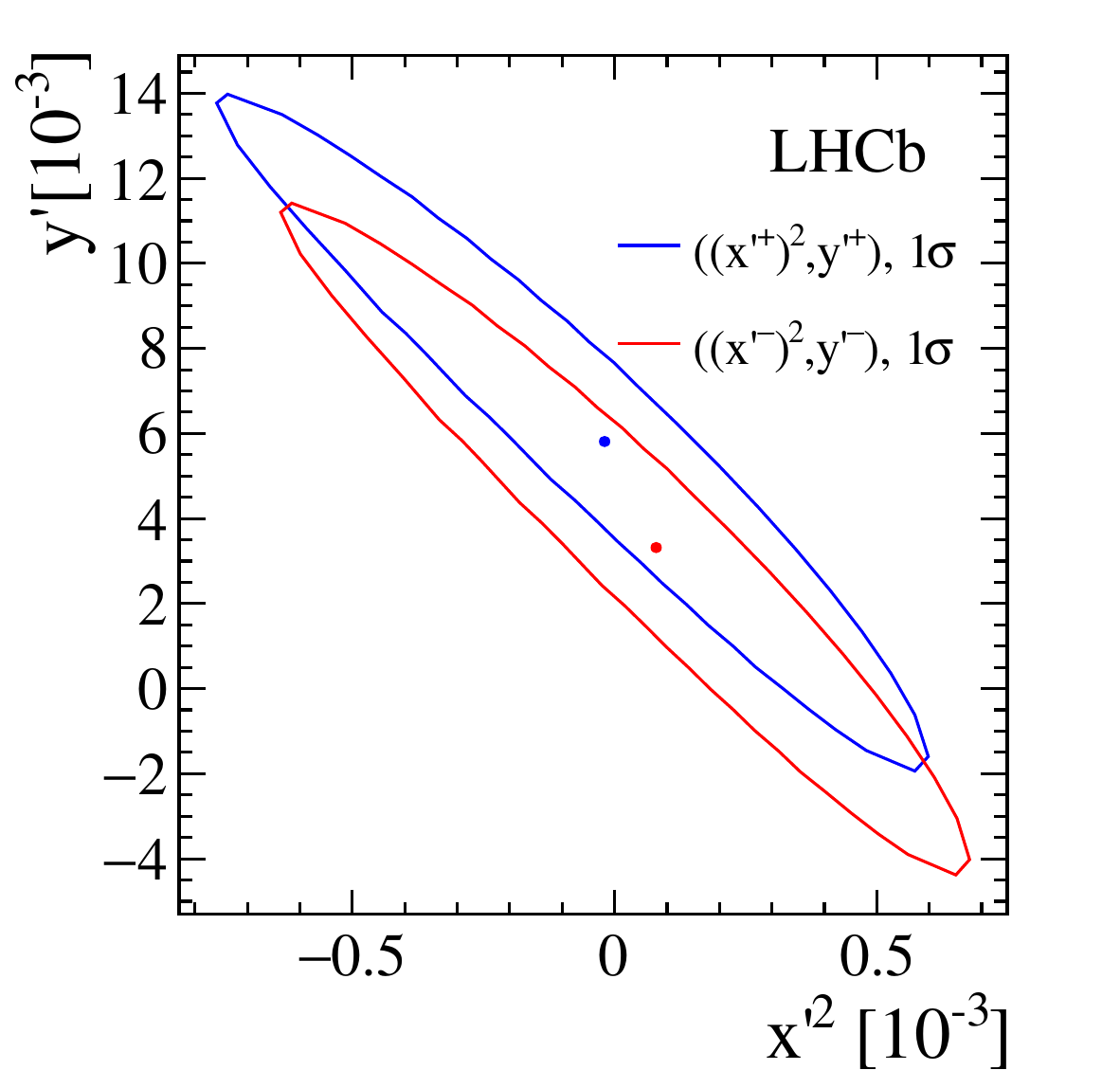}
}
\caption{Contours in $y'$ vs $x'^2$ for the No CPV fit (left), the no direct CPV fit (center),
and for the all CPV allowed fit (right) of the DT only result. For each sample, the dot represents the central value. The 1$\sigma$ contours are shown excluding systematic uncertainties. Blue corresponds to results from \Dz and red to \Dzb.}
\label{fig:theconts}
\end{center}
\end{figure}

\begin{figure}[htbp]
\begin{center}
\resizebox{\textwidth}{!}{
\includegraphics[width=\textwidth]{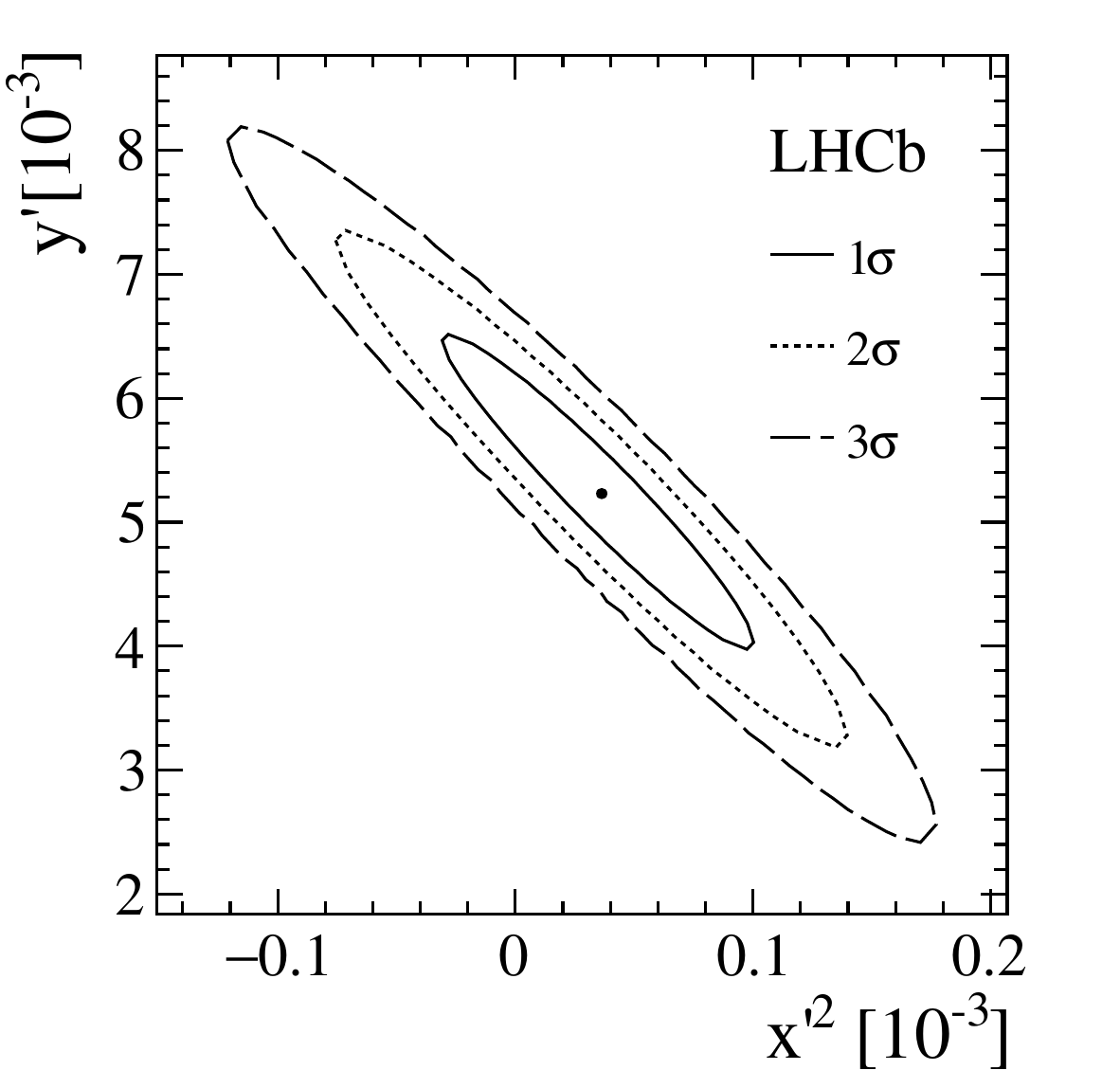}
\includegraphics[width=\textwidth]{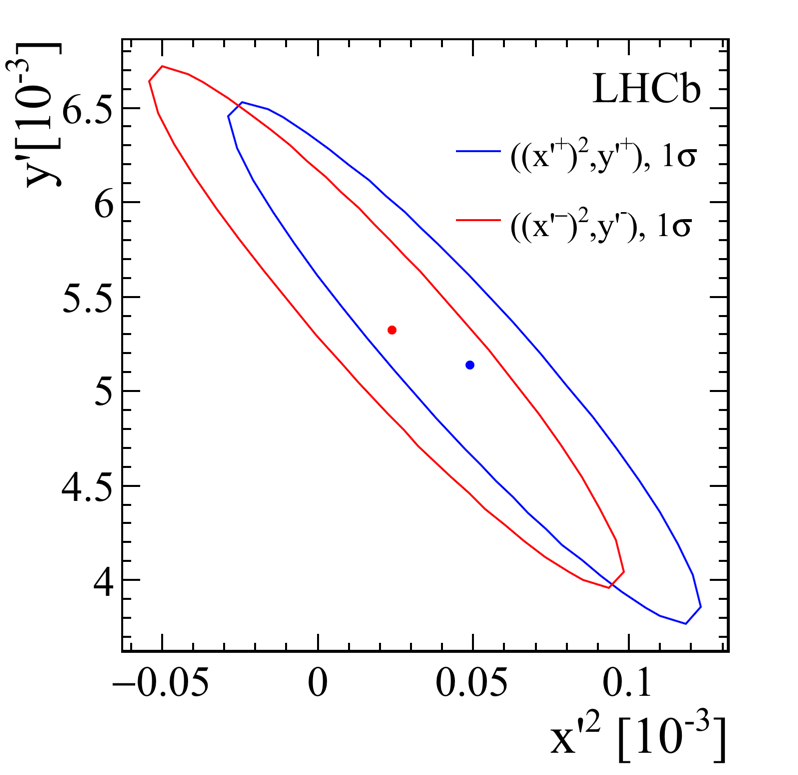}
\includegraphics[width=\textwidth]{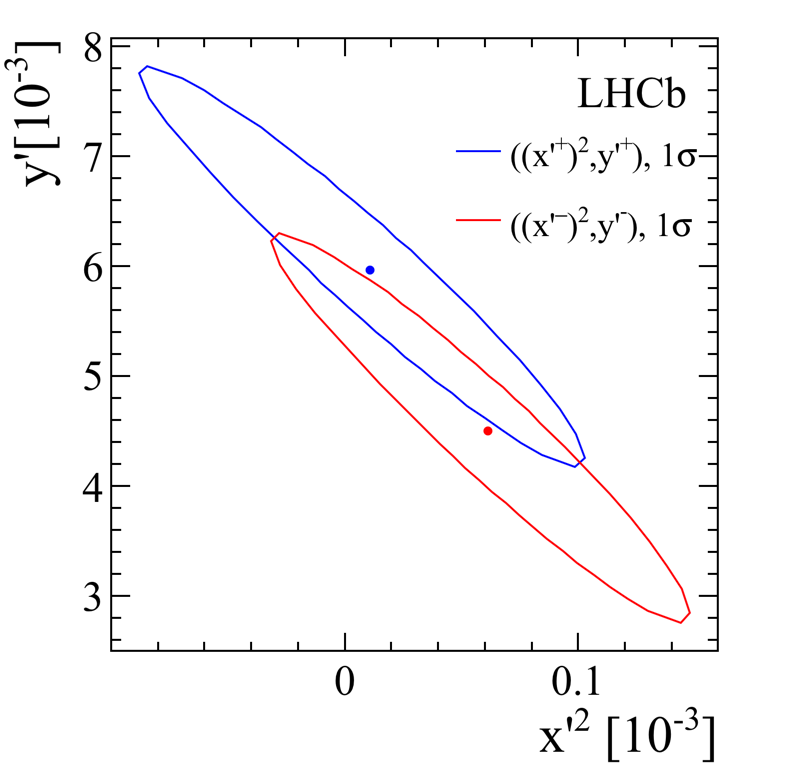}
}
\caption{Contours in $y'$ vs $x'^2$ for the No CPV fit (left), the no direct CPV fit (center),
and for the all CPV allowed fit (right) of the DT + prompt result. For each sample, the dot represents the central value. The 1$\sigma$ contours are shown excluding systematic uncertainties. Blue corresponds to results from \Dz and red to \Dzb.}
\label{fig:theconts}
\end{center}
\end{figure}

\begin{figure}[htbp]
\begin{center}
\resizebox{0.8\textwidth}{!}{
\includegraphics[width=\textwidth]{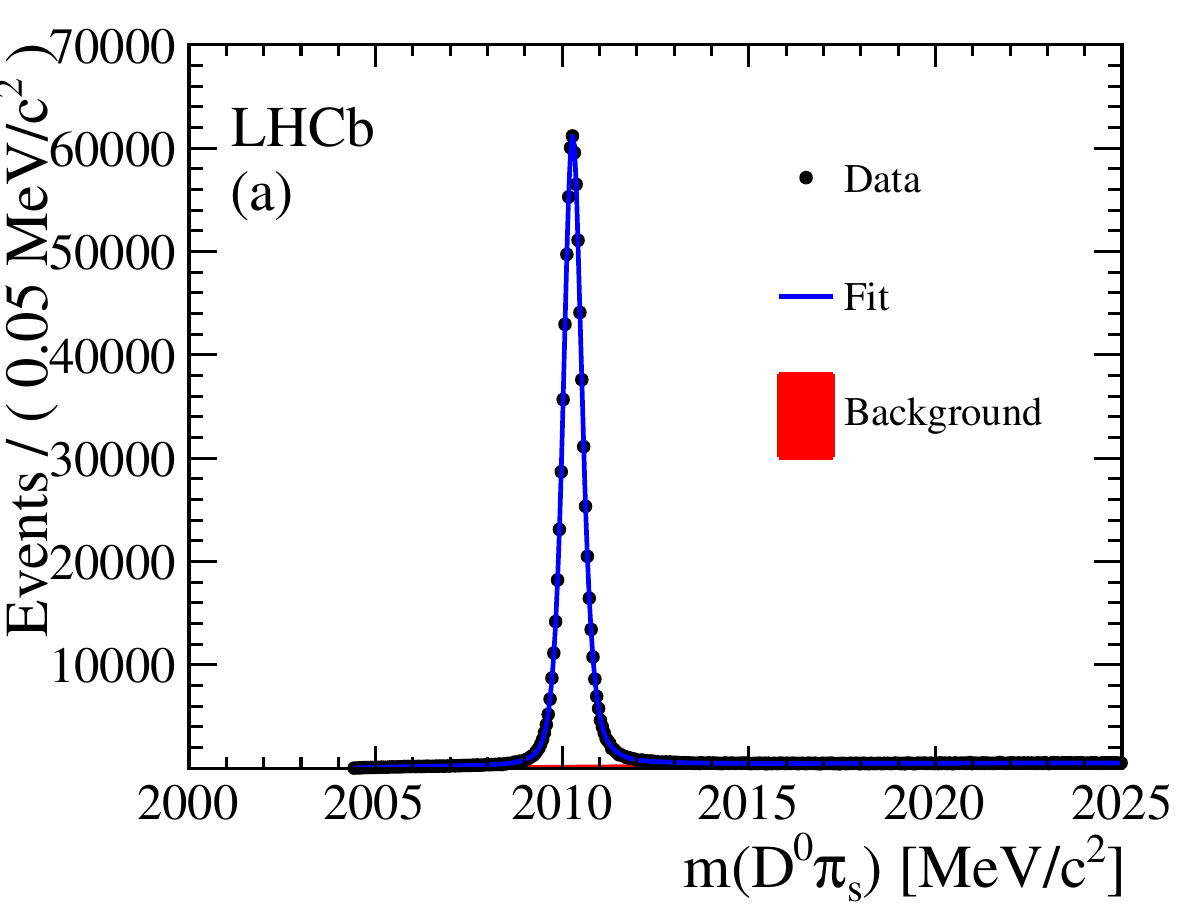}
\includegraphics[width=\textwidth]{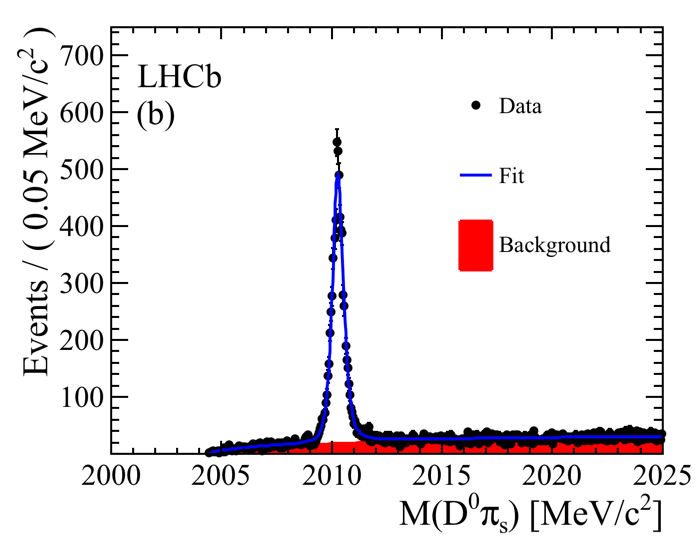}
}
\caption{Fits for the time-integrated RS(a) and WS(b) samples.}
\label{fig:fits_nopull}
\end{center}
\end{figure}

\begin{figure}[htbp]
\begin{center}
\resizebox{0.8\textwidth}{!}{
\includegraphics[width=\textwidth]{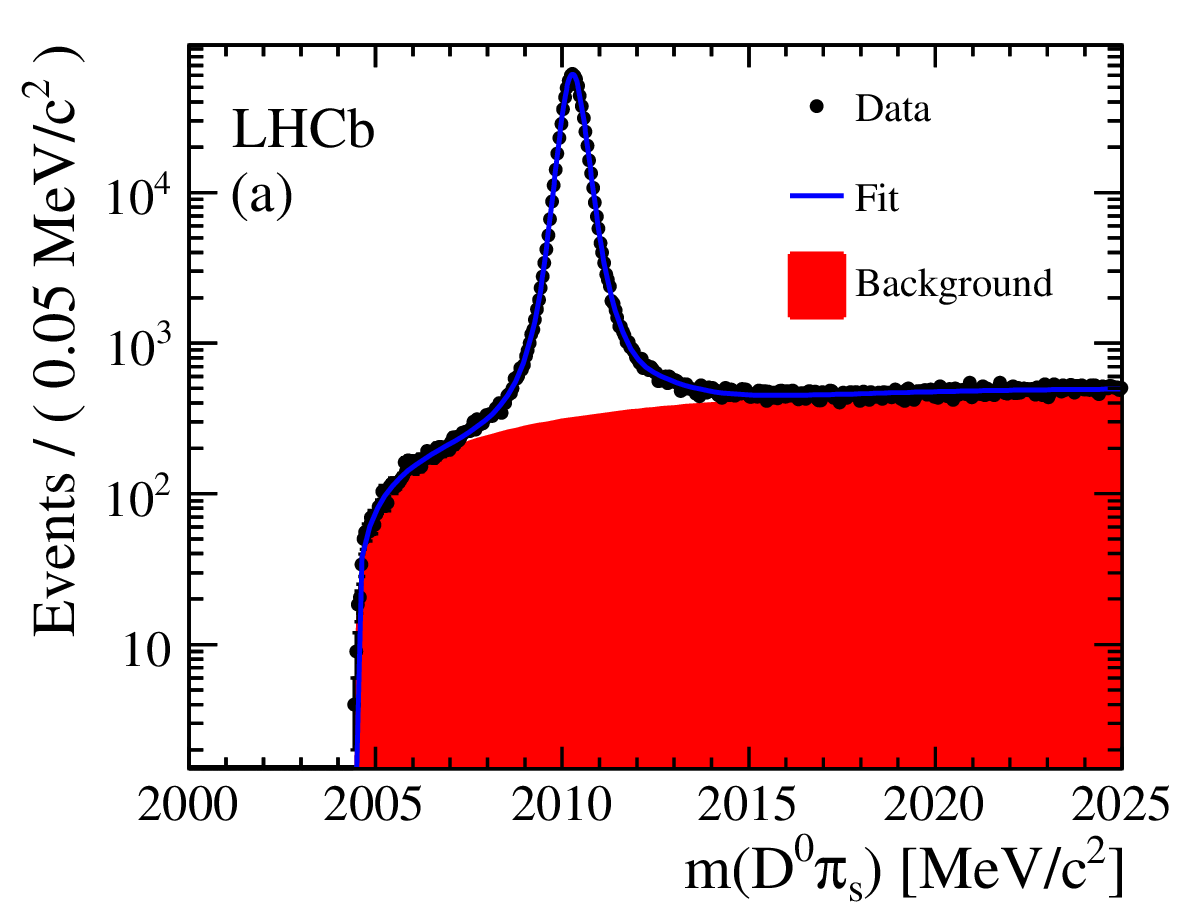}
\includegraphics[width=\textwidth]{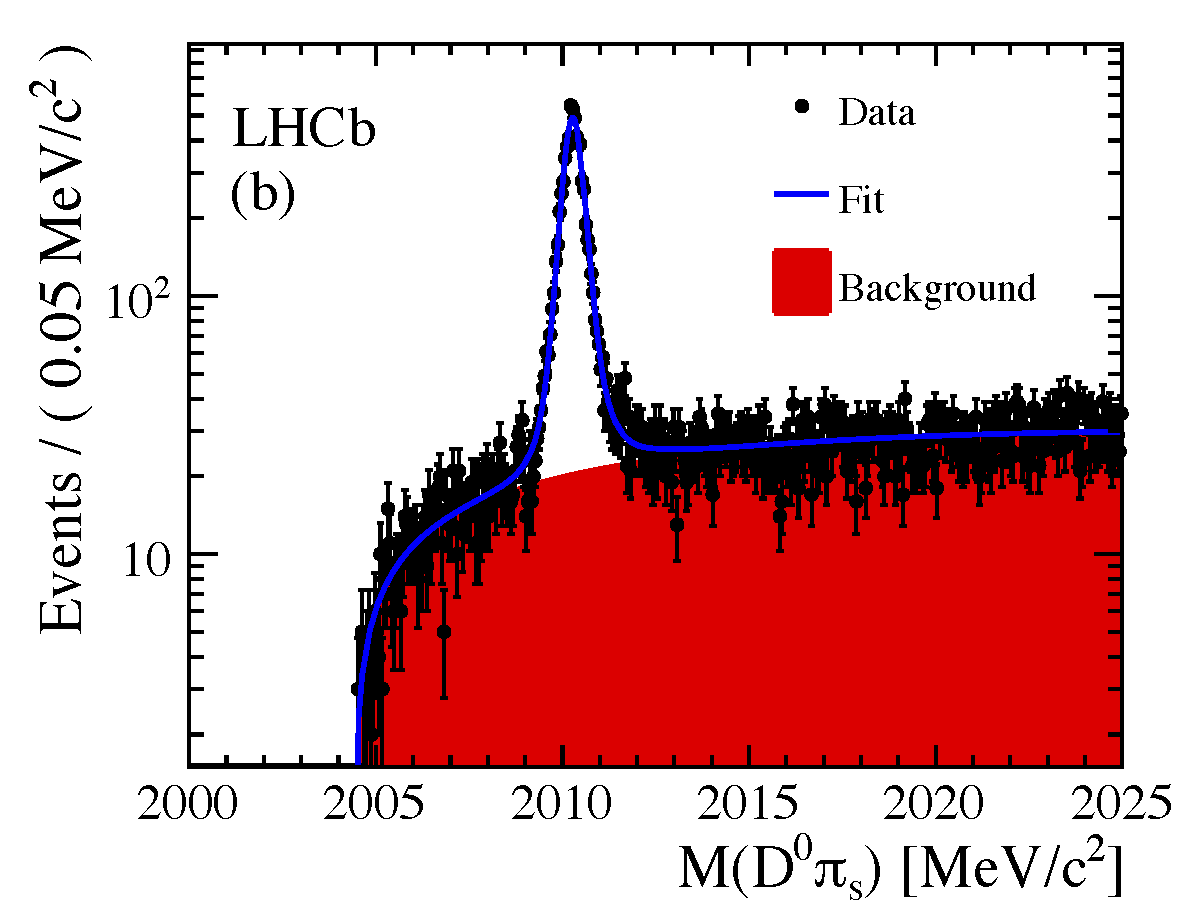}
}
\caption{Fits for the time-integrated RS(a) and WS(b) samples on logarithmic $y$ scale.}
\label{fig:fits_nopull}
\end{center}
\end{figure}

\begin{figure}[htbp]
\begin{center}
\resizebox{0.9\textwidth}{!}{
\includegraphics[width=\textwidth]{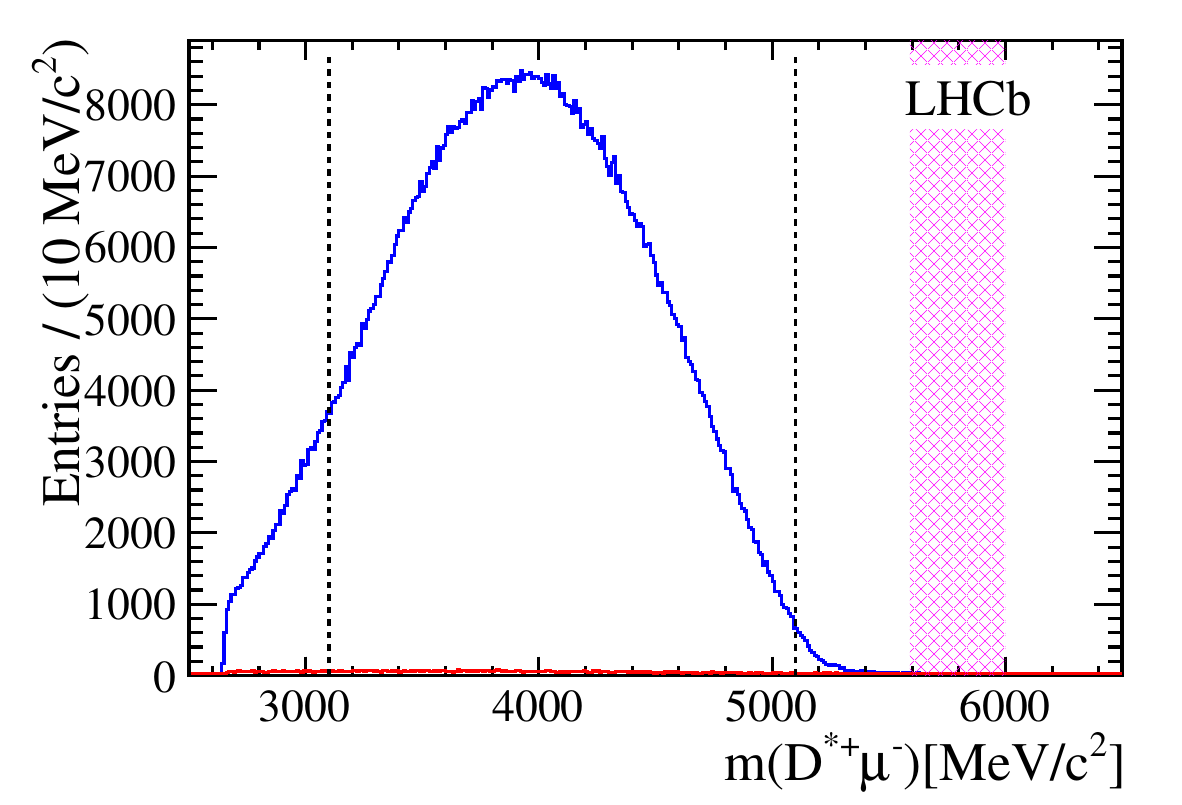}
\includegraphics[width=\textwidth]{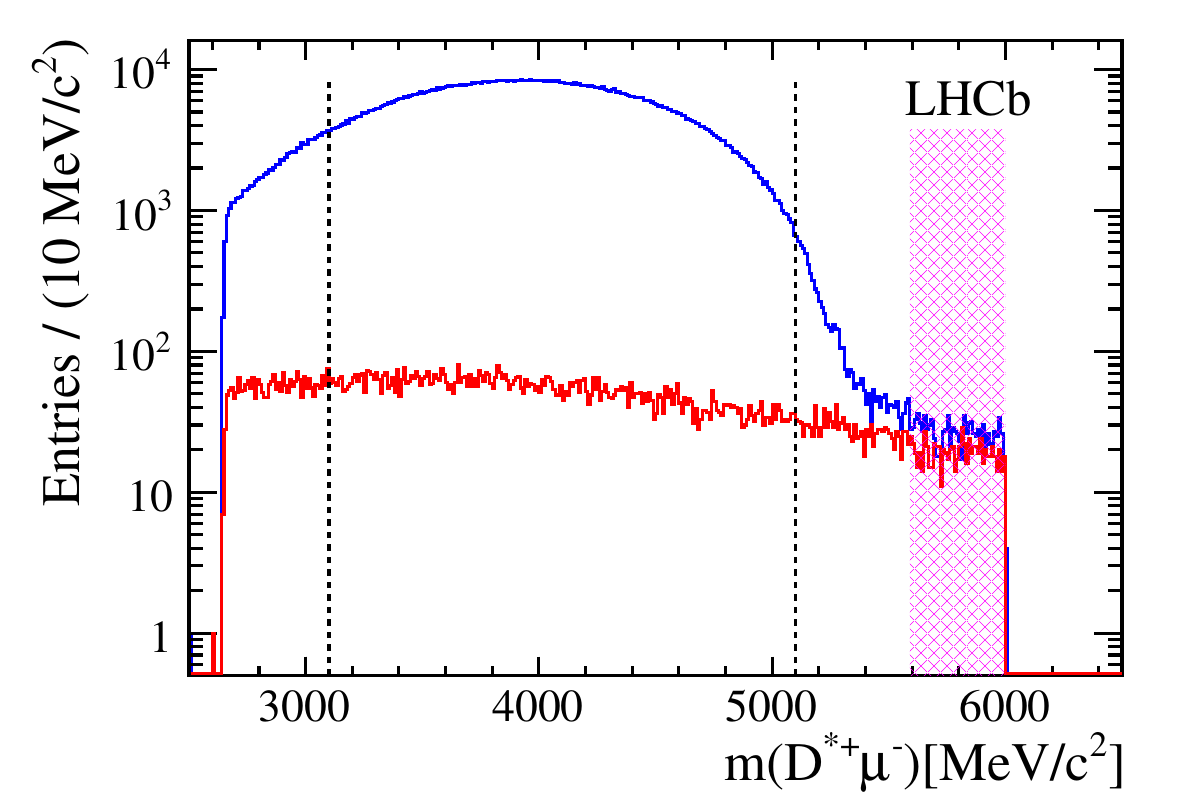}}
\caption{Invariant mass distribution of the $\Dstarp\mu$ combination, integrated over \Dz $t/\tau$. The opposite-sign signal is shown in blue and the same-sign background in red. Overlaid
are the dashed lines, which represent the signal region and the magenta hatched region which represents the scaling region for the same-sign background.}
\label{fig:ss_os_time_integrated_app}
\end{center}
\end{figure}

\begin{figure}[htbp]
\begin{center}
\includegraphics[width=0.4\textwidth]{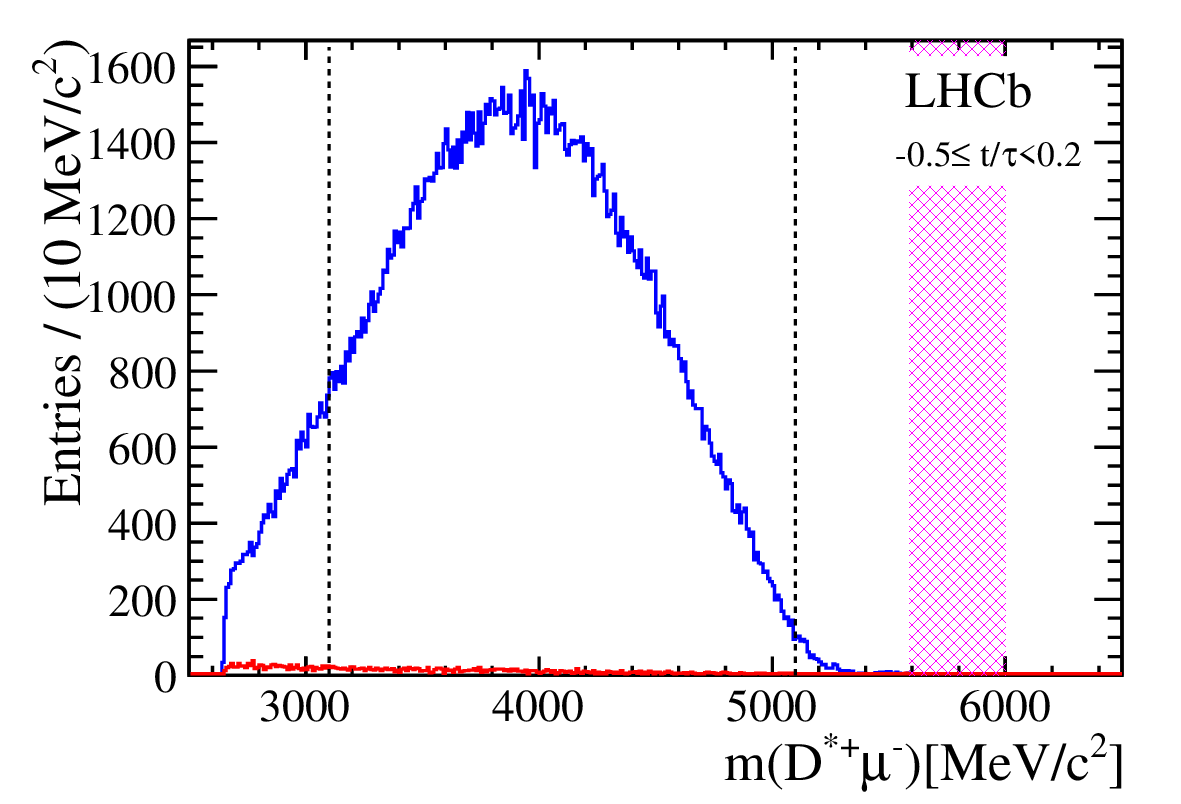}
\includegraphics[width=0.4\textwidth]{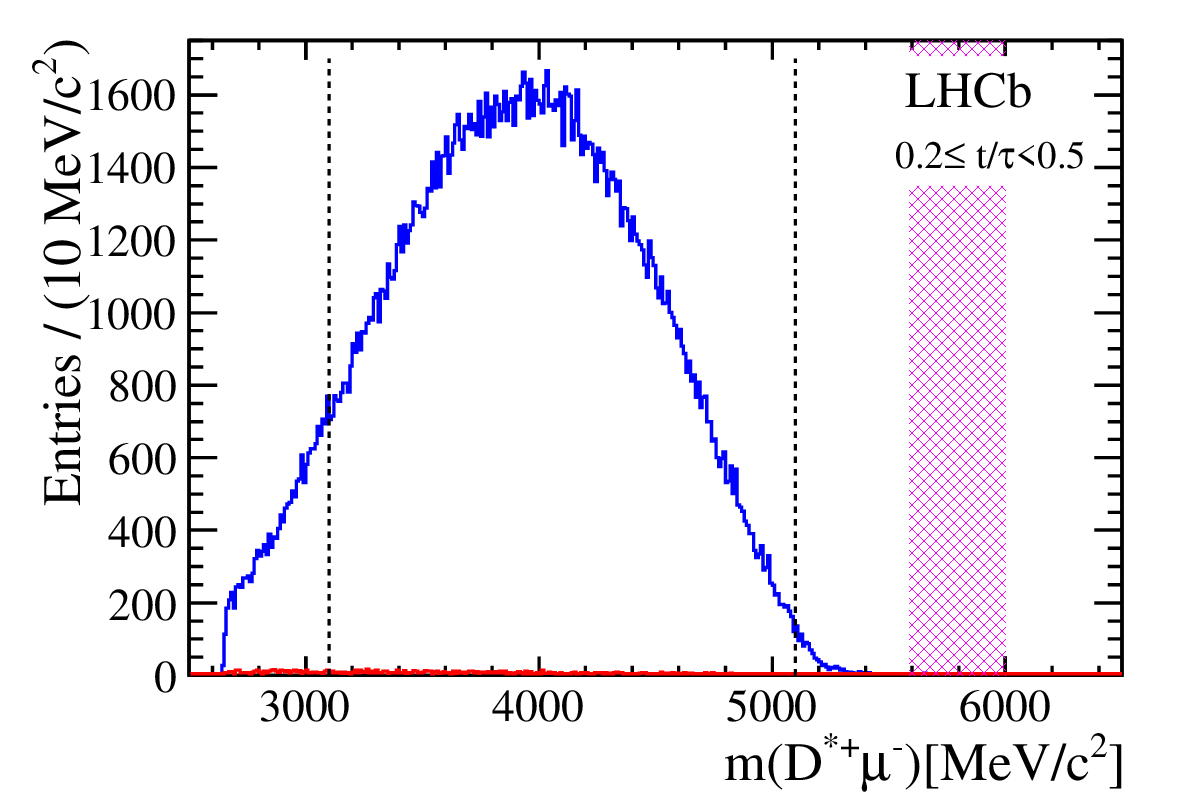}\\
\includegraphics[width=0.4\textwidth]{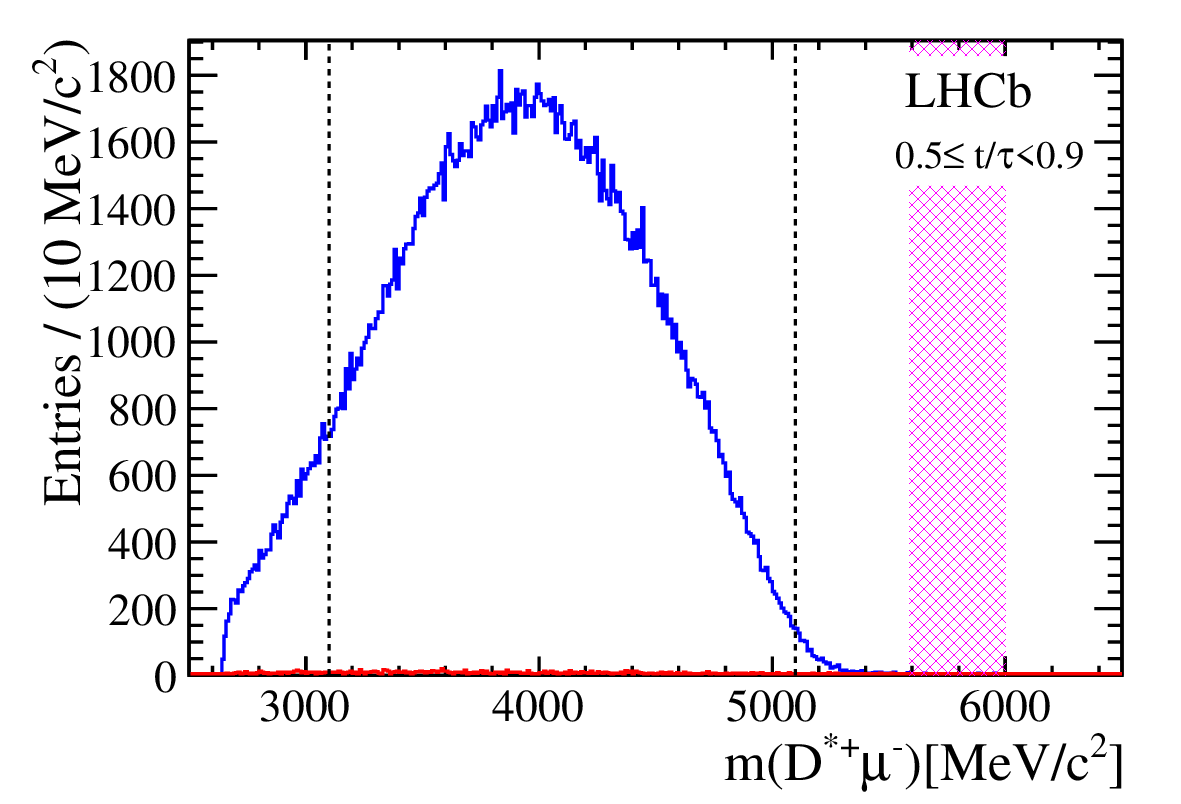}
\includegraphics[width=0.4\textwidth]{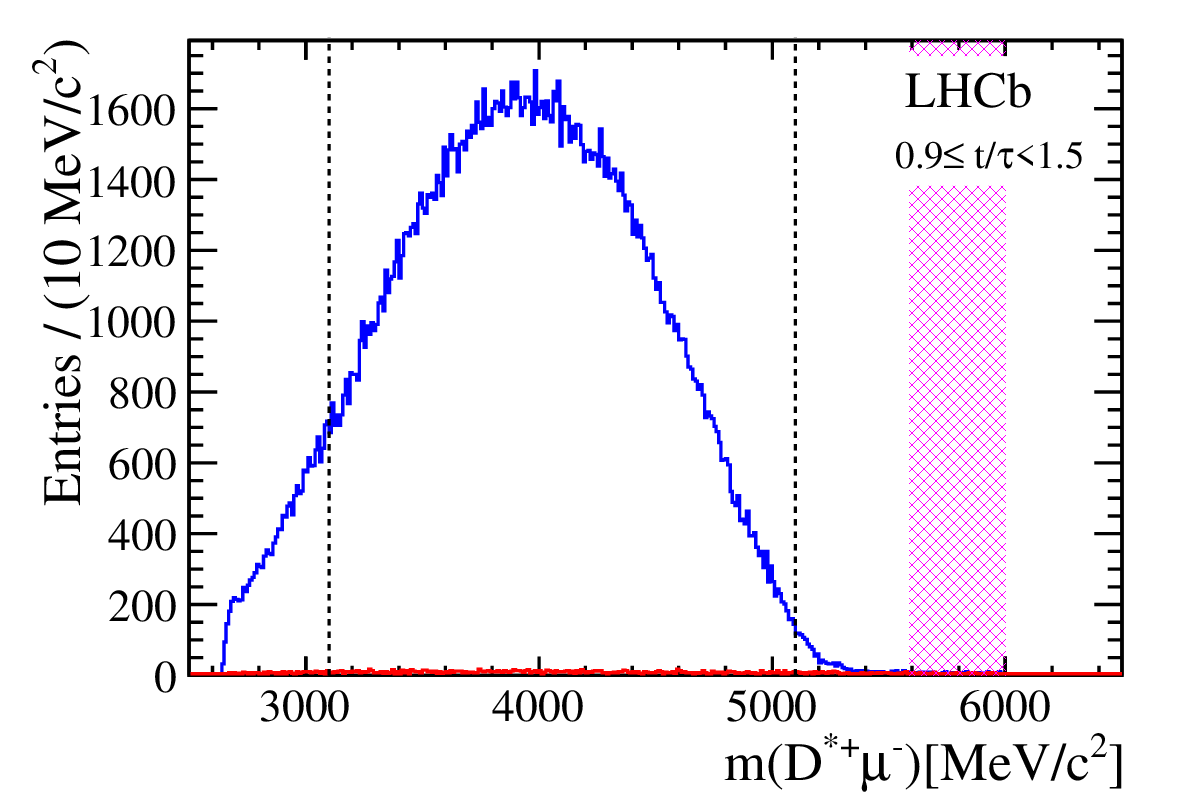}\\
\includegraphics[width=0.4\textwidth]{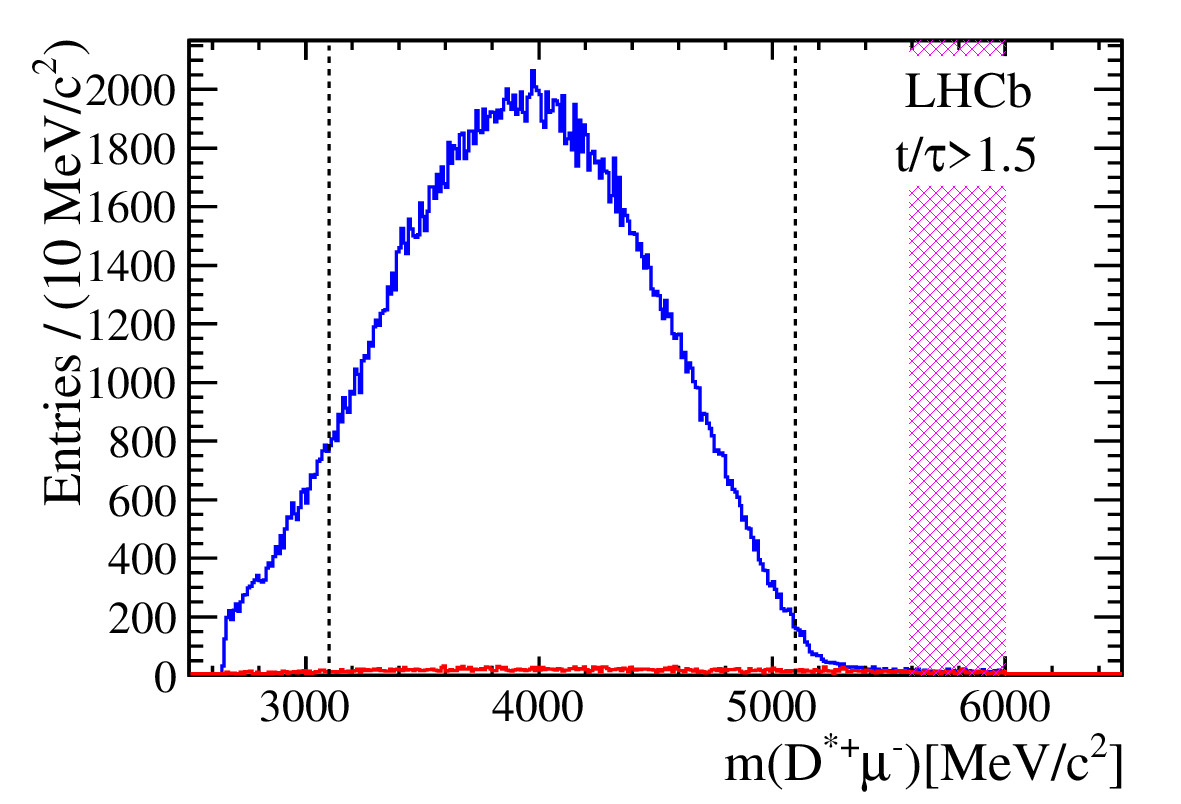}
\caption{Invariant mass distribution of the $\Dstarp \mun $ combination, in bins of \Dz $t/\tau$. The bins used are those in the analysis, 
from the lowest on the top left,
to the highest on the bottom. The opposite-sign signal is shown in blue and the same-sign background in red. Overlaid
are the dashed lines, which represent the signal region and the magenta hatched region which represents the scaling region for the same-sign background.}
\label{fig:ss_os_time_dep_app}
\end{center}
\end{figure}

\begin{figure}[htbp]
\begin{center}
\includegraphics[width=0.4\textwidth]{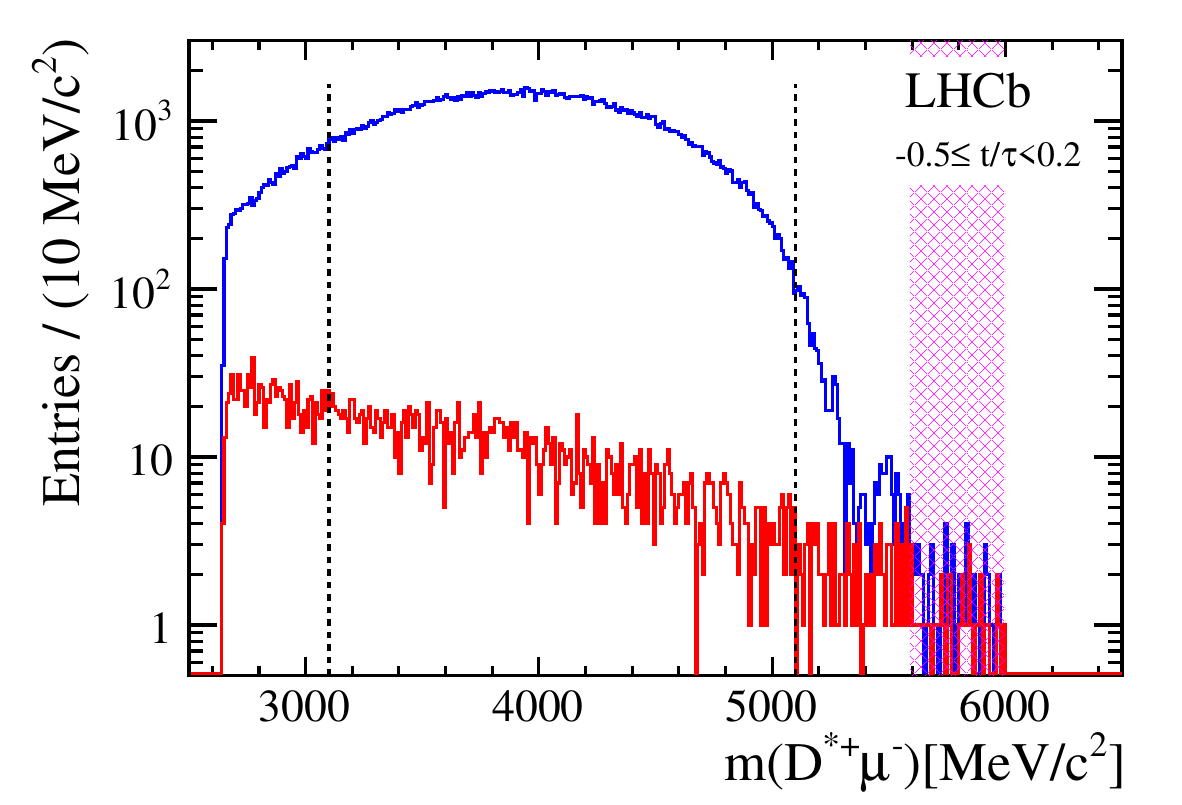}
\includegraphics[width=0.4\textwidth]{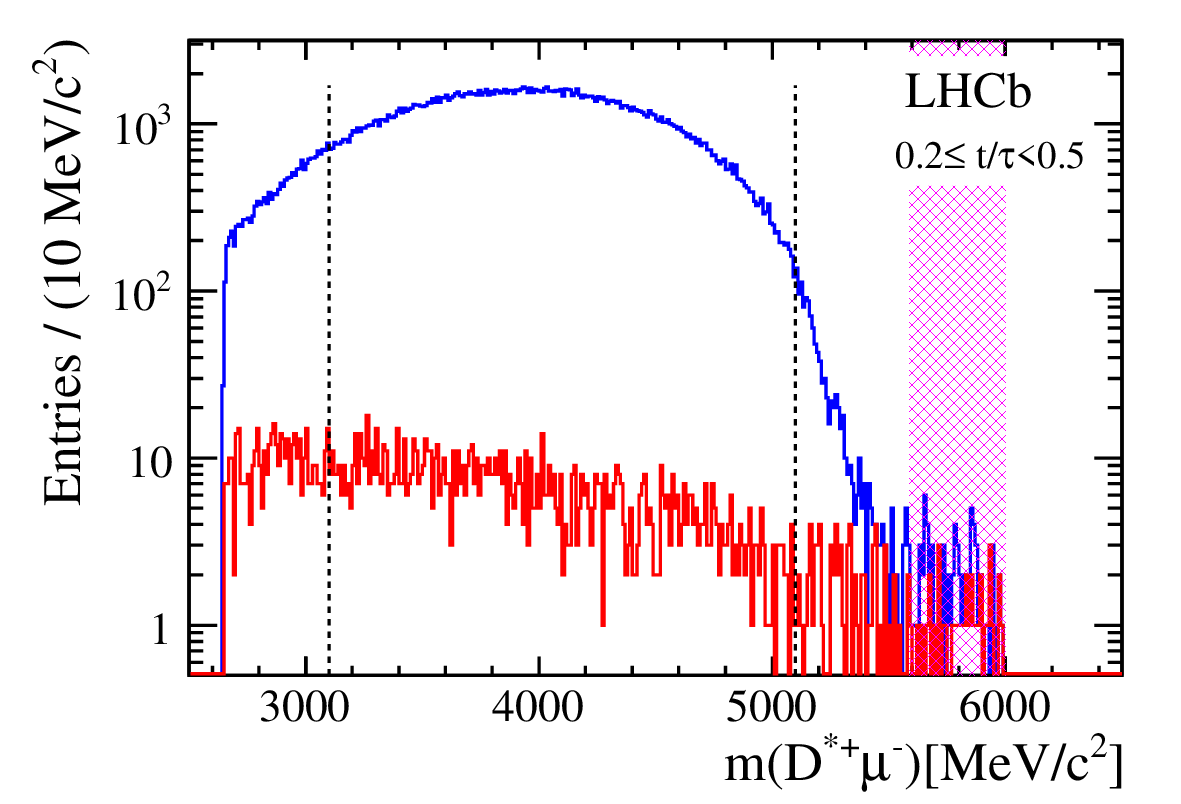}\\
\includegraphics[width=0.4\textwidth]{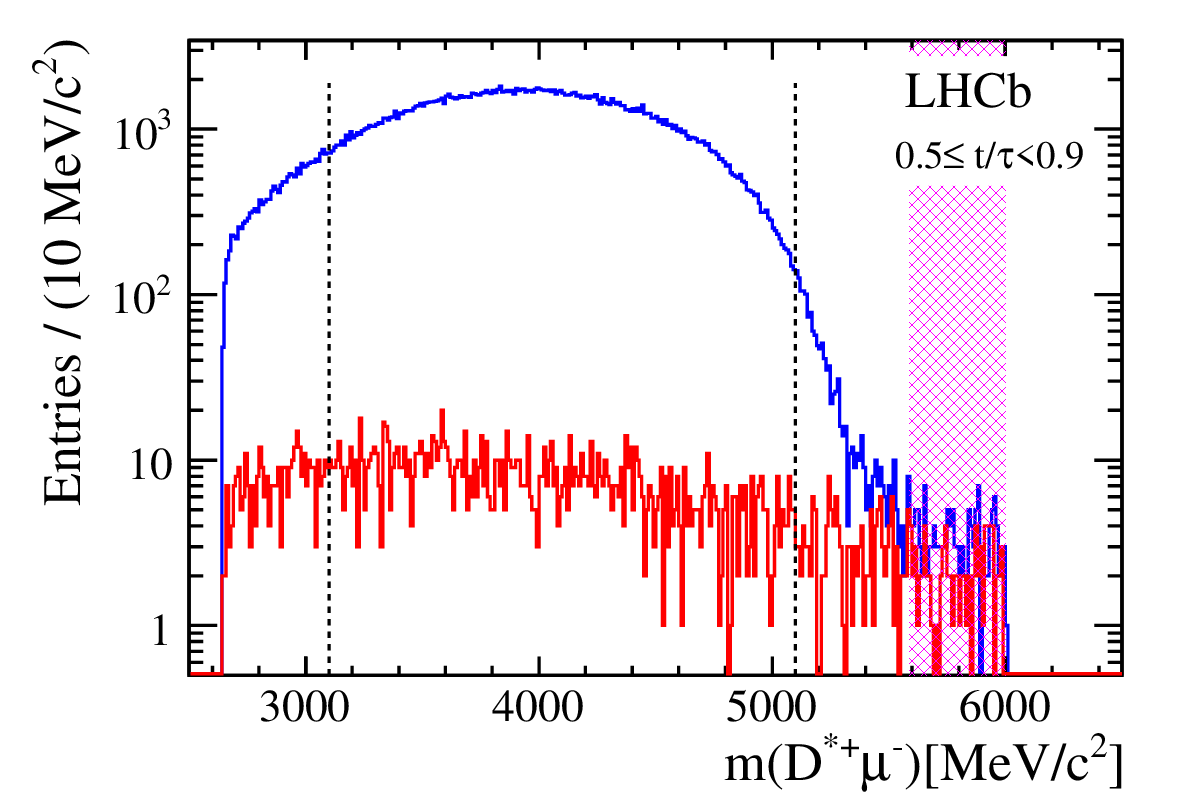}
\includegraphics[width=0.4\textwidth]{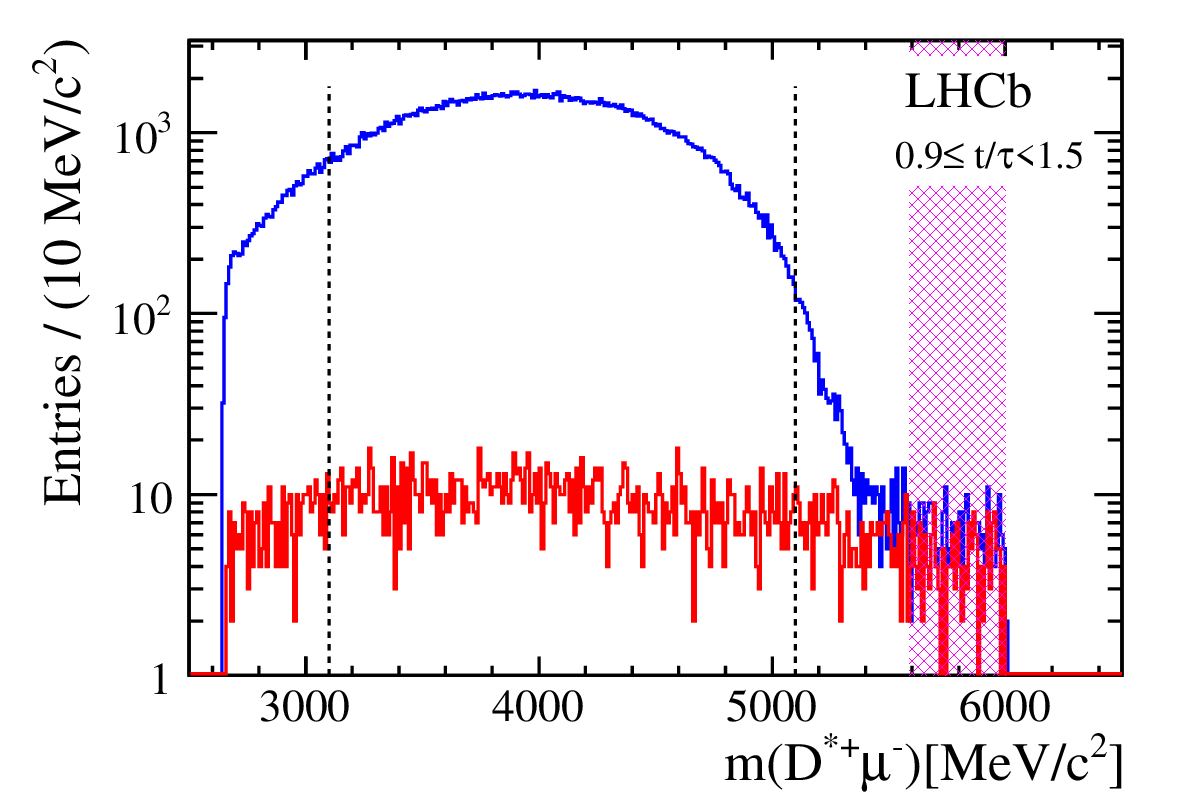}\\
\includegraphics[width=0.4\textwidth]{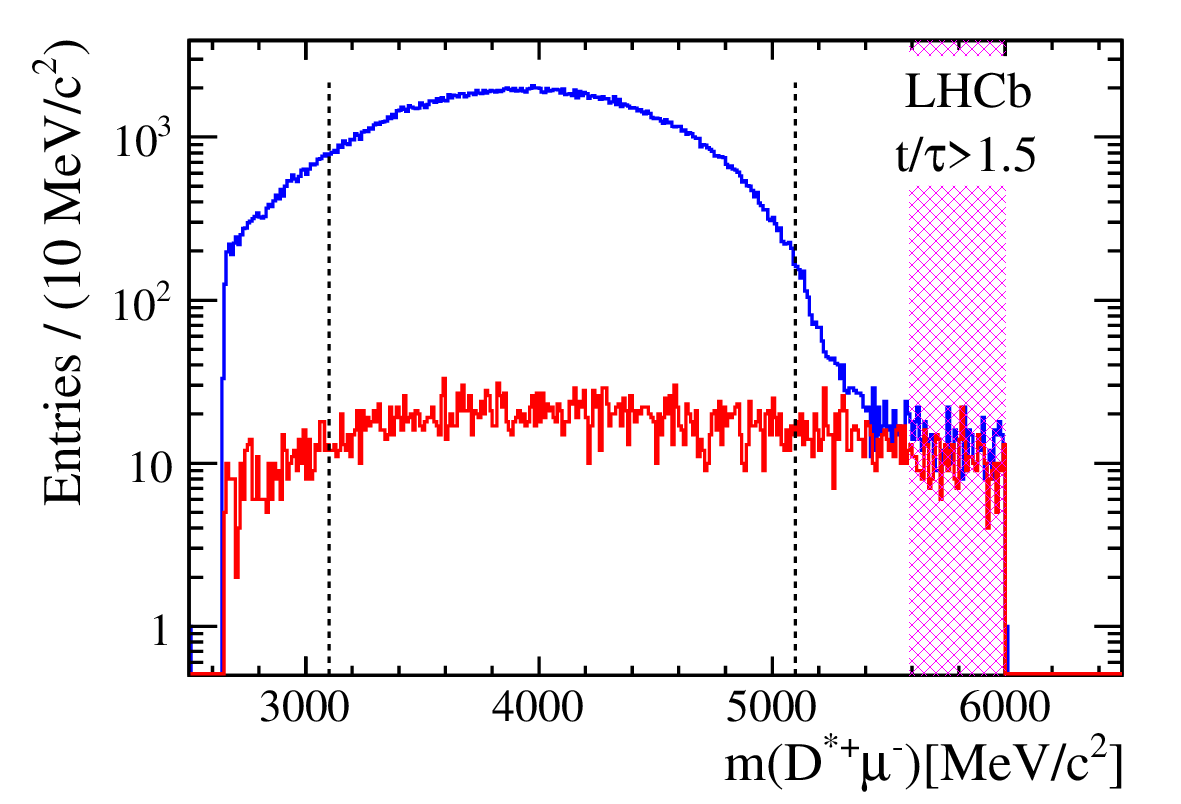}
\caption{Invariant mass distribution of the $\Dstarp\mu $ combination, in bins of \Dz $t/\tau$ on a logarithmic $y$ scale. The bins used are those in the analysis, from the lowest on the top left,
to the highest on the bottom. The opposite-sign signal is shown in blue and the same-sign background in red. Overlaid
are the dashed lines, which represent the signal region and the magenta hatched region which represents the scaling region for the same-sign background.}
\label{fig:ss_os_time_dep_logy_app}
\end{center}
\end{figure}